\documentclass[11pt]{article}
\usepackage{amsfonts}
\usepackage{amsmath}
\usepackage{amssymb}
\usepackage[english]{babel}
\usepackage{color}
\usepackage{geometry}
\usepackage{graphics}
\usepackage{graphicx}
\usepackage[colorlinks=true,urlcolor=blue,linkcolor=blue,citecolor=blue]{hyperref}
\usepackage{lscape}
\usepackage{natbib}
\usepackage{times}

\newcommand{\vzero}{\mathbf{0}}

\newcommand{\va}{\mathbf{a}}
\newcommand{\ve}{\mathbf{e}}
\newcommand{\vh}{\mathbf{h}}
\newcommand{\vm}{\mathbf{m}}
\newcommand{\vn}{\mathbf{n}}
\newcommand{\vv}{\mathbf{v}}
\newcommand{\vx}{\mathbf{x}}
\newcommand{\vy}{\mathbf{y}}

\newcommand{\mA}{\mathbf{A}}
\newcommand{\mB}{\mathbf{B}}
\newcommand{\mC}{\mathbf{C}}

\newcommand{\mG}{\mathbf{G}}
\newcommand{\mH}{\mathbf{H}}
\newcommand{\mI}{\mathbf{I}}
\newcommand{\mQ}{\mathbf{Q}}
\newcommand{\mR}{\mathbf{R}}
\newcommand{\mX}{\mathbf{X}}
\newcommand{\mY}{\mathbf{Y}}
\newcommand{\mZ}{\mathbf{Z}}

\newcommand{\vbeta}{\text{\boldmath{$\beta$}}}
\newcommand{\veps}{\text{\boldmath{$\epsilon$}}}
\newcommand{\veta}{\text{\boldmath{$\eta$}}}
\newcommand{\vmu}{\text{\boldmath{$\mu$}}}
\newcommand{\vtheta}{\text{\boldmath{$\theta$}}}

\newcommand{\mDelta}{\mathbf{\Delta}}
\newcommand{\mSigma}{\mathbf{\Sigma}}
\newcommand{\mOmega}{\mathbf{\Omega}}
\newcommand{\mPsi}{\mathbf{\Psi}}
\newcommand{\mPhi}{\mathbf{\Phi}}

\newcommand{\G}{\mathcal{G}}
\newcommand{\N}{\mathcal{N}}
\newcommand{\U}{\mathcal{U}}
\newcommand{\W}{\mathcal{W}}
\newcommand{\GB}{\mathcal{GB}\mathit{2}}

\newcommand{\argmax}{\mathop{\rm arg~max}\limits}

\title{
	\textbf{Bayesian State-Space Modeling and Model-Based Counterfactual Analysis of Dynamic Income Distributions from Grouped Data}
	\thanks{
		Earlier versions of this paper were circulated under the titles ``Bayesian Analysis of Aging and Declining Household Size Effects on Income Distribution in Japan'' and ``Demographic Transition and the Dynamics of Income Distribution in Japan: A Bayesian State-Space Approach''.
		Previous versions of this paper were presented at the 18th International Joint Conference on Computational and Financial Econometrics (CFE 2024) in London and the 8th International Conference on Econometrics and Statistics (EcoSta 2025) in Tokyo as well as the seminars at Hokkaido University, Kanazawa Gakuin University, and the University of Osaka.
		We would like to thank the seminar/conference participants, especially Yoshihide Kakizawa, Hideo Kozumi, Haruhisa Nishino, Yasuhiro Omori, and Kosuke Oya for their valuable comments and suggestions.
		This work is partially supported by KAKENHI \#20H00080, \#20K01590, and \#26K00322.
	}
}
\author{
	Kazuhiko Kakamu
	\thanks{
		Graduate School of Data Science, Nagoya City University, Yamanohata 1, Mizuho-cho, Mizuho-ku, Nagoya 467-8501, Japan.
		Email: \texttt{\href{mailto:kakamu@ds.nagoya-cu.ac.jp}{kakamu@ds.nagoya-cu.ac.jp}}
	}
}

\date{%
}
\begin{document}
\maketitle
\begin{abstract}
	Grouped income data contain only limited information about the evolution of income distributions over time.
	This paper develops a Bayesian state-space model for the generalized beta distribution of the second kind (GB2) to estimate dynamic income distributions using repeated grouped income data.
	By borrowing information across adjacent periods through the latent GB2 parameters, the proposed framework improves estimation precision relative to independent cross-sectional estimation.
	Building on the estimated latent-state dynamics, we further construct a model-based counterfactual framework that quantifies the contribution of demographic covariates while preserving the estimated evolution of the remaining model components.

	Using Japanese household income data from 1969--2007, we find that population aging and declining household size affect different parts of the income distribution through distinct channels, with population aging becoming an increasingly important driver of income inequality after around 2000.
	More generally, the proposed framework provides a unified Bayesian approach to dynamic distributional analysis and model-based counterfactual inference using repeated grouped income data.

	\vspace*{8pt}

	\vspace*{8pt}

\end{abstract}

\clearpage
\section{Introduction}\label{sec:intro}

Official statistics frequently report income distributions only in grouped form because access to individual-level microdata is restricted by confidentiality or unavailable for long historical periods.
Recovering the evolution of the underlying income distribution from such repeated grouped observations therefore constitutes an important econometric problem.
Because each cross-section contains only limited distributional information, exploiting temporal dependence is essential for obtaining stable estimates of distributional dynamics.

Although a substantial body of literature has examined income inequality and Lorenz curves within parametric frameworks, most studies have focused on cross-sectional or static settings.
Only a limited number of studies have modeled the dynamics of income distributions explicitly.
The dynamics of parametric income distributions were first considered by \citet{NKO12,NK15}, who introduced time-varying parameters into income distribution models.
Subsequently, \citet{KYKKS22} proposed a Bayesian state-space approach for modeling the temporal evolution of Lorenz curves.
More recently, \citet{HHIS24} developed a flexible dynamic framework that estimates Lorenz curves without imposing a specific parametric income distribution by employing basis-function representations.
Taken together, these studies demonstrate that explicitly modeling temporal dependence substantially improves the estimation of evolving income distributions and inequality measures.

Despite these advances, existing approaches primarily describe the evolution of income distributions or summary inequality measures rather than identifying the factors driving those dynamics.
Although \citet{FK22} incorporated income inequality into a vector autoregressive framework to evaluate the effects of monetary policy, no existing study has developed a unified framework that simultaneously estimates the evolution of the entire income distribution and quantifies the contribution of structural determinants to its dynamics.

To fill this gap, we develop a Bayesian state-space model for the generalized beta distribution of the second kind (GB2) using repeated grouped income data.
The proposed framework combines flexible distributional modeling with dynamic Bayesian estimation, allowing information to be borrowed across adjacent periods while simultaneously linking the evolution of the latent income distribution to observable demographic factors.
Building on the estimated latent-state system, we further construct model-based counterfactual income distributions that quantify the contribution of individual demographic factors while preserving the estimated dynamics of the remaining components of the model.

We illustrate the proposed methodology using Japanese household income data previously analyzed by \citet{NKO12,NK15}.
Japan provides a particularly informative empirical setting because rapid population aging and declining household size have long been recognized as important drivers of distributional change (\citealp{T05,O08,Y12}).
More generally, demographic forces, including lifecycle and cohort effects, have long been regarded as fundamental determinants of income inequality (\citealp{DP94}).
Nevertheless, previous studies have largely relied on descriptive statistics, decomposition methods, reduced-form analyses, or structural macroeconomic models.
By contrast, our framework jointly estimates the evolution of the entire income distribution and quantifies how demographic factors reshape different parts of the distribution within a unified Bayesian framework.

Beyond the literature on dynamic income distributions, our study is also related to the literature on distributional counterfactuals and decomposition methods, including the seminal reweighting approach of \citet{DFL96}, the comprehensive survey of \citet{FLF11}, and the econometric framework for counterfactual distributions developed by \citet{CFM13}.
More recently, \citet{M25} extended distributional counterfactual analysis to high-dimensional settings.
Unlike these reduced-form approaches, our framework generates model-based counterfactual income distributions by modifying the contribution of selected demographic covariates within the estimated latent-state equation while preserving all remaining components of the estimated Bayesian state-space model.

This paper makes three main contributions.
First, we develop a Bayesian state-space framework for estimating dynamic income distributions from grouped data, extending existing GB2-based approaches by explicitly exploiting temporal dependence.
Second, the proposed framework borrows information across adjacent periods to substantially improve estimation stability under grouped observations within a coherent Bayesian framework.
Third, we develop a model-based counterfactual framework that quantifies how demographic factors reshape the entire income distribution and aggregate inequality while preserving the estimated latent-state dynamics.
Taken together, these contributions provide a unified Bayesian framework for dynamic distributional analysis and model-based counterfactual inference using repeated grouped income data.

Our empirical analysis shows that population aging and declining household size affect the Japanese income distribution through distinct distributional channels.
Population aging primarily induces persistent changes in the shape of the income distribution, whereas declining household size mainly generates short- and medium-run distributional fluctuations.
The counterfactual analysis further reveals that the contribution of population aging to the Gini coefficient became increasingly pronounced after around 2000, while the effect of declining household size on aggregate inequality remained comparatively modest.
These findings illustrate that modeling the entire income distribution provides insights into demographic change that cannot be obtained from aggregate inequality measures alone.
These results illustrate that analyzing the entire income distribution, rather than relying solely on summary inequality measures, provides substantially richer evidence on how demographic change reshapes economic inequality over time.

The remainder of the paper is organized as follows.
Section \ref{sec:model} introduces the proposed Bayesian state-space model.
Section \ref{sec:posterior} describes posterior inference.
Section \ref{sec:emp} presents the empirical application and model-based counterfactual analysis.
Section \ref{sec:conclusion} concludes.

\section{The Model}\label{sec:model}
The objective of the proposed model is to estimate dynamic income distributions from repeated grouped observations.
Because each cross section contains only limited information about the underlying income distribution, independent estimation may lead to unstable inference.
To address this issue, we introduce a Bayesian state-space specification that exploits temporal dependence in the evolution of GB2 parameters.
By borrowing information across adjacent periods, the proposed model improves estimation stability while providing a coherent framework for model-based counterfactual analysis.

We model the evolution of the GB2 parameters using the following Bayesian state-space specification.
Let $y_{it}$ be the income of $i$th household at time $t$ and $\vx_{t}$ be the $d \times 1$ vector of covariates, which consists of demographic covariates.
The macroeconomic variables are aging rate and the average household size in this study.
We assume that the income follows the generalized beta distribution of the second kind (GB2 distribution) proposed by \citet{M84}.

The GB2 distribution has four parameters ($a, b, p, q$), and its probability density function (PDF) and cumulative density function (CDF) are written as
\begin{align}
	f(x|\vtheta) &= \dfrac{ax^{ap-1}}{b^{ap}B(p,q)\left[ 1+\left( \dfrac{x}{b} \right)^{a} \right]^{p+q}},
	\label{eqn:gb2-pdf}\\
	F(x|\vtheta) &= I_{z}(p,q),\quad \text{with }z = \dfrac{\left( \dfrac{x}{b} \right)^{a}}{1 + \left( \dfrac{x}{b} \right)^{a}},
	\label{eqn:gb2-cdf}
\end{align}
where $\vtheta = (a,b,p,q)^{\prime}$, $B(p,q)$ is a complete beta function and $I_{z}(p,q)$ is a ratio of incomplete and complete beta function.

\begin{figure}[tbp]
	\centering
	\includegraphics[width=\linewidth]{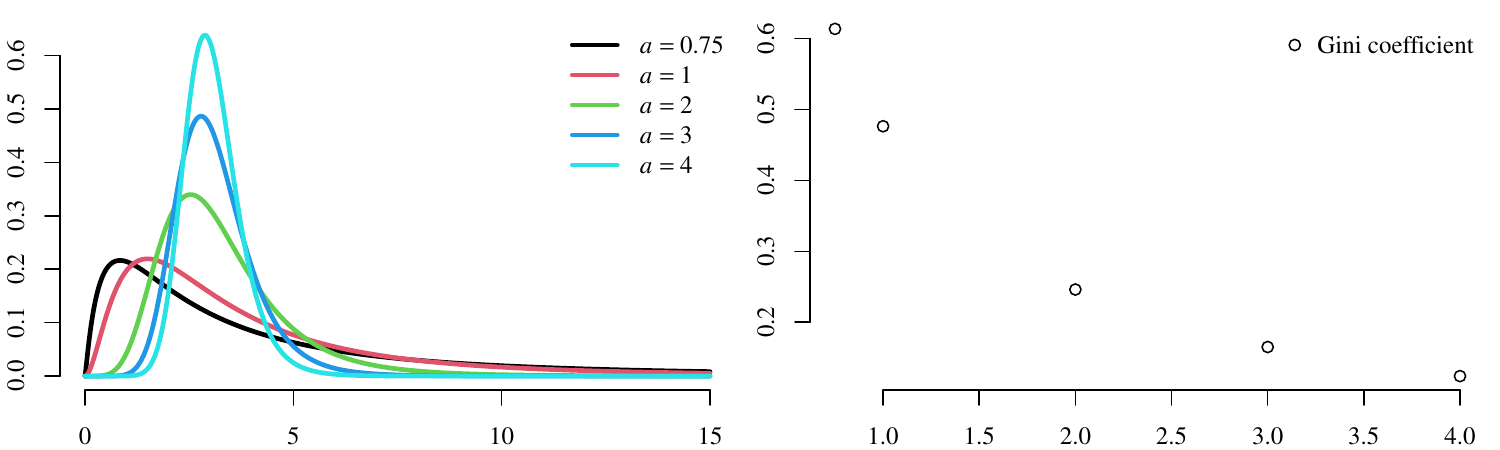}
	\includegraphics[width=\linewidth]{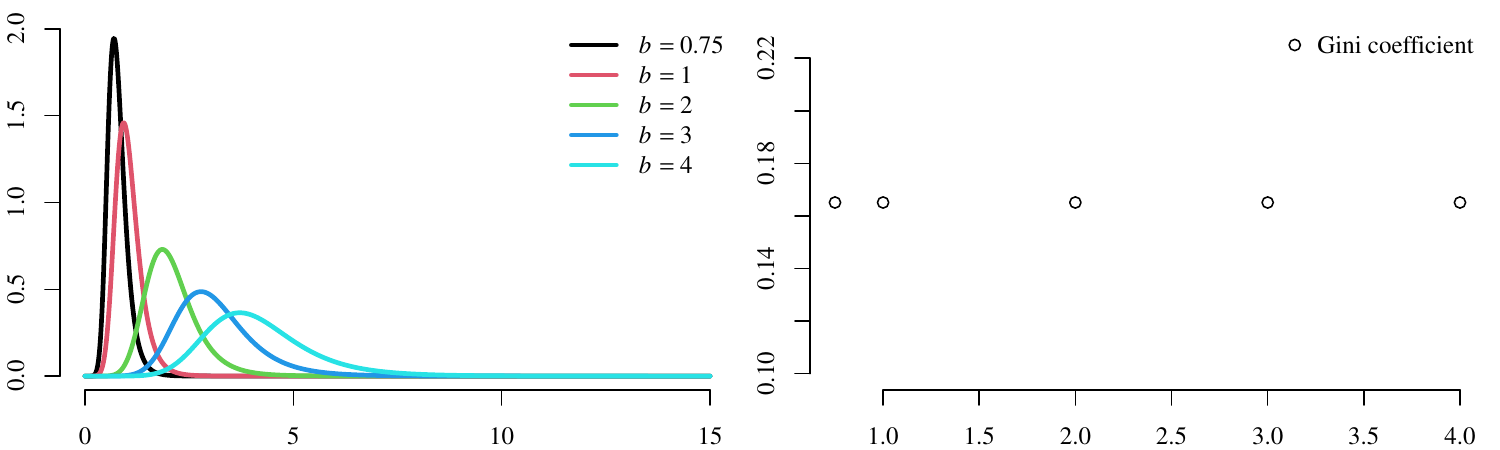}
	\includegraphics[width=\linewidth]{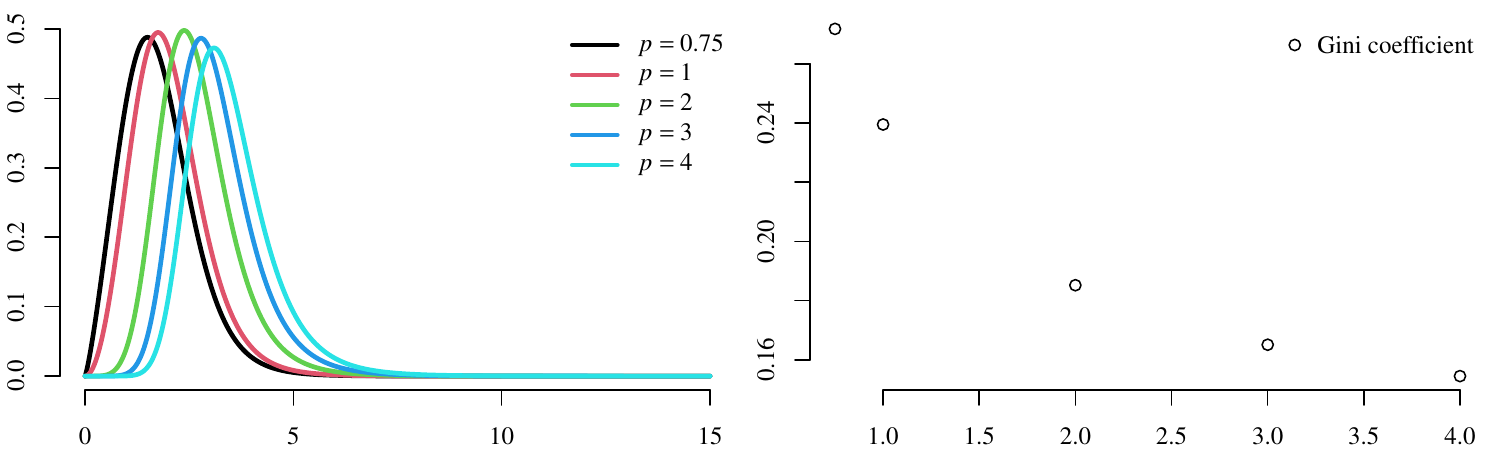}
	\includegraphics[width=\linewidth]{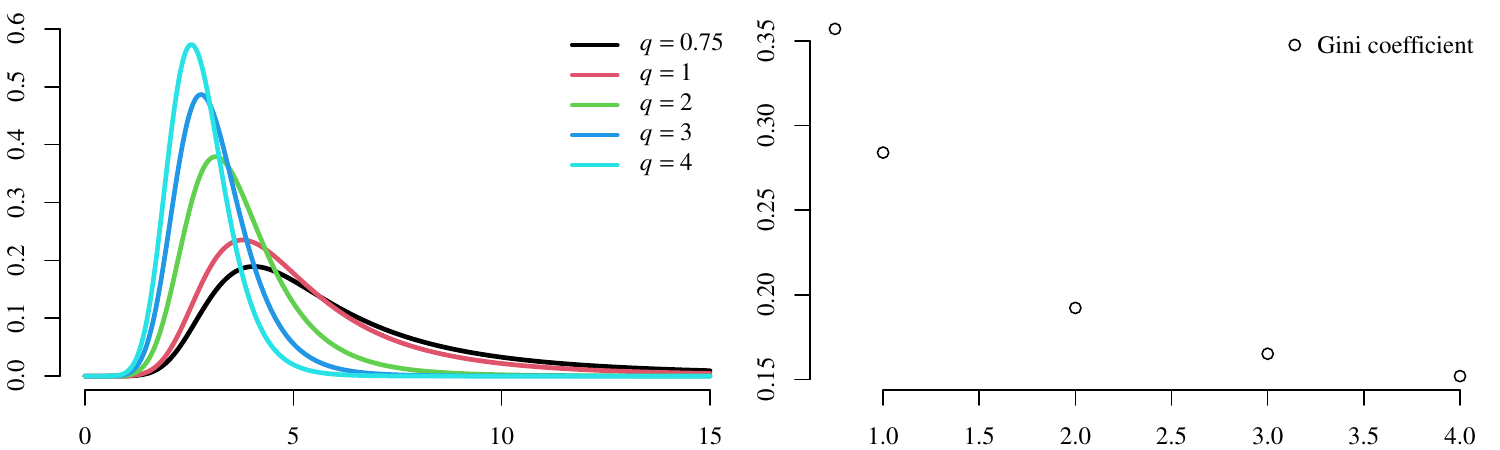}
	\caption{The PDFs of GB2 distribution (left) and their corresponding Gini coefficients (right)}
	\label{fig:GB2}
\end{figure}

\begin{center}[INCLUDE \autoref{fig:GB2} HERE]\end{center}

\autoref{fig:GB2} shows the relationship between the parameters of the GB2 distribution, the corresponding probability density functions, and the associated Gini coefficients.
Changes in the GB2 parameters affect different parts of the income distribution differently, implying that similar Gini coefficients may correspond to substantially different distributional shapes (see \citealp{J09}). This feature is particularly important for understanding how demographic changes reshape the lower and upper tails of the income distribution.
From top to bottom, the figures illustrate changes in $a$, $b$, $p$, and $q$, respectively, while the remaining parameters are fixed at $3$.
From the figure, increasing $a$ reduces the inequality of the distribution while thinning the upper and lower tails of the distribution simultaneously.
In other words, the distribution becomes more concentrated around the mode.
On the other hand, increasing $p$ ($q$) reduces the inequality of the distribution while thinning only the upper (lower) tail of the distribution, respectively.
Different from $a$, $p$ and $q$, $b$ does not change the income inequality, although the shape of the distribution changes.
Therefore, if we can identify which parameters have changed, we can identify how the shape of the income distribution has changed.
Therefore, we examine the time-varying parameters, that is, $\exp(\vh_{t}) = \vtheta_{t} = (a_{t},b_{t},p_{t},q_{t})^{\prime}$.

The GB2 distribution provides a flexible parametric representation of the entire income distribution, with each parameter governing a distinct distributional feature.
Embedding these parameters in a state-space system allows the model to borrow information across adjacent periods, thereby stabilizing estimation under grouped observations.
Moreover, because demographic covariates enter the latent-state equation directly, the framework naturally facilitates model-based counterfactual analysis.

Then, the model is given as follows:
\begin{align}
	y_{it} &\sim \GB(a_{t},b_{t},p_{t},q_{t}),\quad i=1,2,\ldots,n,\quad t=1,2,\ldots,T,\label{eqn-model-obs}\\
	\vh_{t} &= \vmu + \mZ_{t} \vbeta_{t} + \veps_{t},\quad \veps_{t} \sim \N(\vzero, \mOmega),\quad t=1,2,\ldots,T,\label{eqn-model-state1}\\
	\vbeta_{t+1} &= \vbeta_{t} + \veta_{t},\quad \veta_{t} \sim \N(\vzero, \mSigma),\quad t=1,2,\ldots,T-1,\label{eqn-model-state2}
\end{align}
where $\vmu$ is a $4 \times 1$ vector, $\mZ_{t} = \mI_{4} \otimes \vx_{t}^{\prime}$, and $\mI_{n}$ is an $n \times n$ unit matrix.
Therefore, $\vbeta_{t}$ is a $4d \times 1$ vector of coefficients, $\mSigma$ is $4d \times 4d$ variance-covariance matrix, and $\mOmega$ is $4 \times 4$ variance-covariance matrix.
For $\vbeta_{1}$, we assume $\vbeta_{1}\sim \N(\vbeta_{0},\mDelta_{0})$.

The feasibility of estimating GB2 parameters from grouped income data has been demonstrated by \citet{KN19}.
Building on this framework, we introduce a Bayesian state-space specification to exploit temporal dependence in the evolution of income distributions. The purpose of introducing the state-space specification is to borrow information across adjacent periods, rather than estimating each year's GB2 parameters independently, thereby improving estimation stability under grouped observations.
As demonstrated by \citet{KYKKS22}, such temporal borrowing can substantially reduce posterior uncertainty.
Following this idea, the proposed model exploits temporal dependence to improve estimation stability under grouped observations.

In equation \eqref{eqn-model-obs}, it is assumed that $n$ observations are sampled from the GB2 distribution at time $t$.
However, the income data is usually reported in the form of the grouped data.
The grouped data partitions the sample space of observations into $K > 1$ non-overlapping intervals of the forms $(y_{[0]}, y_{[1]}]$, $(y_{[1]}, y_{[2]}]$, $\ldots$, $(y_{[K-1]}, y_{[K]})$, where $y_{[0]} = 0$ and $y_{[K]} =\infty$.
Moreover, only the number, $n_{k}$ of observations falling in each interval $(y_{[k-1]},y_{[k]}]$, $k = 1, 2,\ldots, K-1$, can be observed with $\displaystyle \sum_{k = 1}^{K} n_{k} = n$.
In this paper, as we use quintile data ($K=5$), $y_{[k]}$ for $k=1,2,3,4$ varies over time, while $n_{k}$ is constant over $k$ and $t$, that is, $n_{k} = 2000$ for $t=1,2,\ldots,T$.
Therefore, given the data $\vy_{t}=(y_{[1]},y_{[2]}, \ldots, y_{[K-1]})^{\prime}$ and $\vn = (n_{1}, n_{2}, \ldots, n_{K})^{\prime}$ at time $t$, the joint probability distribution based on the selected order statistics by \citet{NK11} for equation \eqref{eqn-model-obs} at time $t$ is as follows:
\begin{align}
	\lefteqn{\ell(\vy_{t}|\vn,\vh_{t})}\nonumber\\
	&= n! \dfrac{F(y_{[1]}|\vtheta_{t})^{n_{1} - 1}}{(n_{1} - 1)!} f(y_{[1]}|\vtheta_{t}) \left[\prod_{k = 2}^{K-1} \dfrac{\left\{ F(y_{[k]}|\vtheta_{t}) - F(y_{[k - 1]}|\vtheta_{t}) \right\}^{n_{k} - 1}}{(n_{k} - 1)!} f(y_{[k]}|\vtheta_{t}) \right] \dfrac{\left\{ 1 - F(y_{[K-1]}|\vtheta_{t}) \right\}^{n_{K}}}{n _{K}!},
	\label{eqn-likelihood-obs}
\end{align}
where equations \eqref{eqn:gb2-pdf} and \eqref{eqn:gb2-cdf} are substituted in equation \eqref{eqn-likelihood-obs}.

In equation \eqref{eqn-model-state1}, by incorporating the covariates and using an appropriate link function, a regression model is constructed to examine the cause of the changes in the parameters $\vtheta_{t}$.
It should be mentioned that a similar approach has already taken by \citet{NKO12}.
They examined the cause of the income inequality in the case of lognormal distribution using demographic covariates like GDP, aging and so on.
However, they concluded that they could not obtain the fact that demographic covariates affected income inequality.
Although \citet{NKO12} modeled income distributions using the lognormal distribution with time-invariant regression effects, this specification may be restrictive when the underlying distribution exhibits changing tail behavior and when the effects of demographic covariates evolve over time.
To accommodate these features, we employ the GB2 distribution, which provides substantially greater flexibility in modeling the shape of income distributions, and allow the regression coefficients, $\vbeta_t$, to evolve over time through the state-space specification.
However, \citet{S26} shows that the FIES tends to report lower measured income inequality than other Japanese surveys.
To account for this persistent survey-specific level difference, we include a vector intercept term, $\vmu$, in the state equation for the latent GB2 parameters.
Then, the probability density function of $\vh_{t}$ at time $t$ is expressed as follows:
\begin{align}
	g(\vh_{t}|\vx_{t},\vmu,\vbeta_{t},\mOmega^{-1}) = \dfrac{1}{\sqrt{2\pi}^{4}}|\mOmega|^{-\dfrac{1}{2}}\exp\left\{ -\frac{(\vh_{t} - \vmu - \mZ_{t}\vbeta_{t})^{\prime}\mOmega^{-1}(\vh_{t} - \vmu - \mZ_{t}\vbeta_{t})}{2} \right\}
	\label{eqn-likelihood-state1}
\end{align}

In equation \eqref{eqn-model-state2}, we assume that the dynamics of $\vbeta_{t}$ follows a random walk process.
However, $\vbeta_{t}$ is not observed.
Therefore, if the $\vbeta_{t}$ for $t=1,\ldots,T$ are observed, the probability density function of $\vbeta_{t+1}$ at time $t + 1$ is expressed as follows:
\begin{align}
	h(\vbeta_{t+1}|\vbeta_{t},\mSigma^{-1}) = \frac{1}{\sqrt{2\pi}^{4d}}|\mSigma|^{-\frac{1}{2}}\exp\left\{ -\frac{(\vbeta_{t+1}-\vbeta_{t})^{\prime}\mSigma^{-1}(\vbeta_{t+1}-\vbeta_{t})}{2} \right\}
	\label{eqn-likelihood-state2}
\end{align}

Given the joint distributions above, the likelihood function is expressed if the $\vbeta_{t}$ for $t=1,2,\ldots,T$ is observed.
Although they are not observed, suppose that they are possible to be augmented easily and consider the augmented likelihood function.
Then, the augmented likelihood function is defined as follows:
\begin{align}
	\lefteqn{L(\mY | \vn, \mX, \mH, \vmu, \mB, \mOmega^{-1}, \mSigma^{-1})}\nonumber\\
	&\quad\quad = \left\{ \prod_{t=1}^{T} \ell(\vy_{t}|\vn, \vh_{t})g(\vh_{t}|\vx_{t},\vmu,\vbeta_{t},\mOmega^{-1}) \right\}h(\vbeta_{1}|\vbeta_{0},\mSigma_{0}^{-1})\left\{ \prod_{t=1}^{T-1}h(\vbeta_{t+1}|\vbeta_{t},\mSigma^{-1}) \right\},
	\label{eqn:likelihood}
\end{align}
where $\mY=(\vy_{1},\vy_{2}\ldots,\vy_{T})$, $\mX=(\vx_{1},\vx_{2},\ldots,\vx_{T})$, $\mH = (\vh_{1},\vh_{2},\ldots,\vh_{T})$, and $\mB = (\vbeta_{1},\ldots,\vbeta_{T})$.

Because we adopt a Bayesian approach, we complete the model by specifying the prior distribution over the parameters.
We apply the following prior:
\begin{align}
	\pi(\vmu,\mOmega^{-1},\mSigma^{-1}) = \pi(\vmu)\pi(\mOmega^{-1})\pi(\mSigma^{-1}).
	\label{eqn-prior}
\end{align}
Given a prior density in \eqref{eqn-prior} and the augmented likelihood function in \eqref{eqn:likelihood}, the joint posterior distribution can be expressed as
\begin{align}
	\pi(\mH,\vmu,\mB,\mOmega^{-1},\mSigma^{-1}|\mY,\vn,\mX) \propto \pi(\vmu,\mOmega^{-1},\mSigma^{-1})L(\mY | \vn, \mX, \mH, \vmu, \mB, \mOmega^{-1}, \mSigma^{-1}).
	\label{eqn:posterior}
\end{align}

Finally, we assume the following prior distributions:
\begin{align*}
	\vmu \sim \N(\vmu_{0}, \mPhi_{0}),\quad \mOmega^{-1} \sim \W(n_{0}, \mOmega_{0}),\quad \mSigma^{-1} \sim \W(m_{0}, \mSigma_{0}),
\end{align*}
where $\W$ represents a Wishart distribution.
\footnote{
	Wishart priors are adopted for computational convenience and because they remain widely used in Bayesian state-space and time-varying parameter (TVP) models.
	Recent work shows that shrinkage priors can substantially improve variance estimation in dynamic settings—for example, the triple-gamma prior of \citet{CFK20}.
	Incorporating such shrinkage priors into the dynamic GB2 framework is a promising direction for future research.
}

The dynamic structure of our model also plays a central role in the counterfactual analysis conducted later in the paper.
Our approach is conceptually related to the literature emphasizing the role of identifying assumptions in model-based counterfactual analysis (\citealp{CC23}), although their focus is on robustness rather than dynamic counterfactual paths, and to recent work on policy counterfactuals in dynamic environments (\citealp{MW23}), which highlights how counterfactual trajectories can be constructed within forward‑looking macroeconomic models.
These insights are particularly relevant in our setting, where counterfactual income distributions are generated through the dynamic evolution of GB2 parameters.

\section{Posterior Analysis}\label{sec:posterior}

To obtain the posterior estimates, we implement the following MCMC steps:
\begin{enumerate}
		\renewcommand{\labelenumi}{Step \theenumi.}
	\item Set $m=1$ and initial values $\mH^{(0)}$, $\vmu^{(0)}$, $\mB^{(0)}$, $\mOmega^{(0)-1}$, and $\mSigma^{(0)-1}$
	\item Draw $\vh_{t}^{(m)}$ from $\pi(\vh_{t}|\vy_{t},\vn,\vx_{t},\vmu^{(m-1)},\vbeta_{t}^{(m-1)}, \mOmega^{(m-1)-1})$ for $t=1,2,\ldots,T$, sequentially.\label{step2}
	\item Draw $\vmu^{(m)}$ from $\pi(\vmu|\mX,\mH^{(m)},\mB^{(m-1)},\mOmega^{(m-1)-1})$.
	\item Draw $\mB^{(m)}$ from $\pi(\mB|\mX,\mH^{(m)},\vmu^{(m)},\mOmega^{(m-1)-1},\mSigma^{(m-1)-1})$.
	\item Draw $\mOmega^{(m)-1}$ from $\pi(\mOmega^{-1}|\mX,\mH^{(m)},\vmu^{(m)},\mB^{(m)})$.
	\item Draw $\mSigma^{(m)-1}$ from $\pi(\mSigma^{-1}|\mB^{(m)})$.
	\item Return to Step \ref{step2}, and set $m$ to $m+1$.
\end{enumerate}
To proceed the above steps, to derive the full conditional distributions from equation \eqref{eqn:likelihood} and appropriate algorithms are required for each step.
The details are given in the next subsections.

\subsection{Sampling \texorpdfstring{$\vh_{t}$}{h\_t} for \texorpdfstring{$t=1,\ldots,T$}{t = 1,..., T}}

The full conditional distribution of $\vh_{t}$ is
\begin{align}
	\pi(\vh_{t} | \vy_{t}, \vn, \vx_{t}, \vmu, \vbeta_{t}, \mOmega^{-1}) \propto \ell(\vy_{t} | \vn, \vh_{t} ) g(\vh_{t}|\vx_{t},\vmu, \vbeta_{t},\mOmega^{-1}).
	\label{eqn:fcd-theta}
\end{align}
From the distribution, It is difficult to find the standard form like a multivariate normal distribution.
Therefore, we need to implement the estimation via a Metropolis--Hastings algorithm.
However, if the sample size is not large enough and/or the number of groups are small, it is difficult to estimate the parameters of the GB2 distribution from grouped data by a random walk Metropolis-Hastings algorithm by \citet{CG00,K16}.
Then, \citet{KN19} showed that a Tailored randomized block Metropolis-Hastings (TaRBMH) algorithm first proposed by \citet{CR10} for estimating the parameters of a DSGE model is required to accelerate the convergence of MCMC draws and to estimate the parameters of the generalized beta distribution efficiently.
Therefore, the TaRBMH algorithm is utilized and the algorithm for the $m$th step is as follows:
\begin{enumerate}
	\item Separate $\vh_{t}$ into $2 \times 1$ vectors $\vh_{t1}^{(m-1)}$ and $\vh_{t2}^{(m-1)}$ randomly.
	\item For $j = 1,\ 2$, implement the following Metropolis--Hastings steps.
		\begin{enumerate}
			\item Generate $\vh_{tj}^{new}$ from a multivariate $t$ distribution, $t(\hat{\vh}_{tj}, \mPsi_{tj}, \nu)$, with mean $\hat{\vh}_{tj}$, covariance $\mPsi_{tj}$, and $\nu$ degrees of freedom.
				Here,
				\begin{equation*}
					\hat{\vh}_{tj} = \argmax_{\vh_{tj}}~\left\{ \log \ell(\vy_{t} | \vn, \vh_{t} ) + \log g(\vh_{t}|\vx_{t},\vmu^{(m-1)},\vbeta_{t}^{(m-1)},\mOmega^{(m-1)-1}) \right\},
				\end{equation*}
				$\vh_{t} = (\vh_{t1}^{\prime}, \vh_{t2}^{(m-1) \prime})^{\prime}$ for $j = 1$, and $\vh_{t} = (\vh_{t1}^{(m) \prime}, \vh_{t2}')'$ for $j = 2$ using simulated annealing.
				Here,
				\begin{equation*}
					\mPsi_{tj} = \left. \left( -\frac{\partial^{2} \left\{ \log \ell(\vy_{t} | \vn, \vh_{t} ) + \log g(\vh_{t}|\vx_{t},\vmu^{(m-1)},\vbeta_{t}^{(m-1)},\mOmega^{(m-1)-1}) \right\}}{\partial \vh_{tj} \partial \vh_{tj}'} \right)^{-1} \right|_{\vh_{tj} = \hat{\vh}_{tj}}.
				\end{equation*}
			\item If $j = 1$, compute
				\begin{align*}
					\lefteqn{\alpha_{1}(\vh_{t1}^{(m-1)}, \vh_{t1}^{new} )}\\
					&= \min \left\{ \frac{\pi(\vh_{t1}^{new} | \vh_{t2}^{(m-1)}, \vy_{t}, \vn, \vx_{t}, \vmu^{(m-1)}, \vbeta_{t}^{(m-1)},\mOmega^{(m-1)-1}) q(\vh_{t1}^{(m-1)} | \hat{\vh}_{t1}, \mPsi_{t1})}{\pi(\vh_{t1}^{(m-1)} | \vh_{t2}^{(m-1)}, \vy_{t}, \vn, \vx_{t}, \vmu^{(m-1)}, \vbeta_{t}^{(m-1)},\mOmega^{(m-1)-1}) q(\vh_{t1}^{new} | \hat{\vh}_{t1}, \mPsi_{t1})}, 1 \right\},
				\end{align*}
				and if $j = 2$, compute
				\begin{align*}
					\lefteqn{\alpha_{2}(\vh_{t2}^{(m-1)}, \vh_{t2}^{new} )}\\
					&= \min \left\{ \frac{\pi(\vh_{t2}^{new} | \vh_{t1}^{(m)}, \vy_{t}, \vn, \vx_{t}, \vmu^{(m-1)}, \vbeta_{t}^{(m-1)},\mOmega^{(m-1)-1}) q(\vh_{t2}^{(m-1)} | \hat{\vh}_{t2}, \mPsi_{t2})}{\pi(\vh_{t2}^{(m-1)} | \vh_{t1}^{(m)}, \vy_{t}, \vn, \vx_{t}, \vmu^{(m-1)}, \vbeta_{t}^{(m-1)},\mOmega^{(m-1)-1}) q(\vh_{t2}^{new} | \hat{\vh}_{t2}, \mPsi_{t2})}, 1 \right\},
				\end{align*}
				where $q(\vh_{tj}^{new} | \hat{\vh}_{tj}, \mPsi_{tj})$ is a multivariate $t$ distribution given in (a).
			\item Generate a value $u_{j}$ from $\U(0,1)$, where $\U(a,b)$ is an uniform distribution on the interval $(a,b)$.
			\item If $u_{j} \le \alpha_{j}( \vh_{tj}^{(m-1)}, \vh_{tj}^{new} )$, set $\vh_{tj}^{(m)} = \vh_{tj}^{new}$, otherwise $\vh_{tj}^{(m)} = \vh_{tj}^{(m-1)}$.
		\end{enumerate}
\end{enumerate}

\subsection{Sampling \texorpdfstring{$\mB$}{\textbf{B}}}

Although \citet{CPS92} present a broadly applicable Bayesian sampling framework for state-space models, in the specific setting considered here the algorithm of \citet{CK94} is preferred due to its greater computational efficiency and exactness.
\citet{CK94} exploit the conditional linear-Gaussian structure of the model to implement forward filtering-backward sampling (FFBS), which draws the entire latent state trajectory conditional on the static parameters in a single, exact pass rather than by iteratively proposing and updating individual state components.
This approach reduces autocorrelation in the state draws, avoids the need for Metropolis-Hastings proposals inside the Gibbs sampler, and yields more reliable mixing for the latent states in practice.
For these reasons---and because our model satisfies the conditional linear-Gaussian assumptions required by the FFBS routine---we adopt the \citet{CK94} sampler for drawing the latent states while retaining the broader \citet{CPS92} framework for sampling the static parameters when appropriate.
This algorithm consists of the following two steps:

\begin{enumerate}
	\item \textbf{Forward Filtering}:

		For $t=1,2,\ldots,T$, recursively compute the filtered distribution
		\begin{align*}
			\vbeta_{t} | \mH_{1 : t}^{(m)} \sim \N(\vm_{t},\mC_{t}),
		\end{align*}
		where $\mH_{1:t}^{(m)} = \left(\vh_{1}^{(m)},\vh_{2}^{(m)},\ldots,\vh_{t}^{(m)}\right)$ and using the Kalman filter equations:
		\begin{align*}
			\va_{t} &= \vm_{t-1},\\
			\mR_{t} &= \mC_{t-1} + \mSigma^{(m-1)},\\
			\mQ_{t} &= \mZ_{t} \mR_{t} \mZ_{t}^{\prime} + \mOmega^{(m-1)},\\
			\mA_{t} &= \mR_{t}\mZ_{t}^{\prime}\mQ_{t}^{-1},\\
			\vm_{t} &= \va_{t} +\mA_{t}\left(\vh_{t}^{(m)} - \vmu^{(m)} -\mZ_{t}\va_{t}\right),\\
			\mC_{t} &= \mR_{t} -\mA_{t} \mZ_{t}\mR_{t},
		\end{align*}
		where $\vm_{0} = \vbeta_{0}$ and $\mC_{0} = \mDelta_{0}$.
		Note that no sampling is performed at this stage; we only store the mean and covariance parameters of the filtering distributions.
	\item \textbf{Backward Sampling}:

		Once the filtering distributions are computed, we sample the latent states $\mB$ in a backward pass.
		First, we sample the terminal state:
		\begin{align}
			\vbeta_{T} \sim \N(\vm_{T}, \mC_{T}),
		\end{align}
		and then for $t=T-1,T-2,\ldots,1$, we sample each $\vbeta_{t}$ from the conditional distribution:
		\begin{align*}
			\vbeta_{t} | \vbeta_{t+1}^{(m)}, \mH^{(m)} \sim \N(\hat{\vm}_{t}, \hat{\mC}_{t}),
		\end{align*}
		where $\hat{\vm}_{t} = \vm_{t} + \mG_{t}\left(\vbeta_{t+1}^{(m)} - \vm_{t}\right)$, $\hat{\mC}_{t} = \mC_{t} - \mG_{t}\left(\mC_{t} + \mSigma^{(m-1)}\right)\mG_{t}^{\prime}$, and $\mG_{t} = \mC_{t}\left(\mC_{t} + \mSigma^{(m-1)}\right)^{-1}$.
\end{enumerate}

\subsection{Sampling the Other Parameters}
The full conditional distribution of $\vmu$ is given by
\begin{align}
	\vmu | \mX, \mH^{(m)}, \mB^{(m-1)}, \mOmega^{-1(m-1)} \sim \N(\hat{\vmu},\hat{\mPhi}),
	\label{eqn:fcd-mu}
\end{align}
where $\hat{\mPhi} = \left( T \mOmega^{-1(m-1)} + \mPhi_{0}^{-1} \right)^{-1}$ and $\hat{\vmu} = \hat{\mPhi}\left\{ \mOmega^{-1(m-1)} \sum_{t=1}^{T} \left( \vh_{t}^{(m)} - \mZ_{t} \vbeta_{t}^{(m-1)} \right) + \mPhi_{0}^{-1} \vmu_{0} \right\}$.

The full conditional distribution of $\mOmega^{-1}$ is given by
\begin{align}
	\mOmega^{-1}| \mX, \mH^{(m)}, \mB^{(m)} \sim \W(\hat{n},\hat{\mOmega}),
	\label{eqn:cdf-omega}
\end{align}
where $\hat{n} = n_{0} + T$, $\hat{\mOmega} = \left( \sum_{t=0}^{T}\ve_{t}\ve_{t}^{\prime} + \mOmega_{0}^{-1}\right)^{-1}$, $\ve_{t} = \vh_{t}^{(m)} - \vmu^{(m)} - \mZ_{t} \vbeta_{t}^{(m)}$.

The full conditional distribution of $\mSigma^{-1}$ is given by
\begin{align}
	\mSigma^{-1}|\mB^{(m)} \sim \W(\hat{m},\hat{\mSigma}),
	\label{eqn:cdf-sigma}
\end{align}
where $\hat{m} = m_{0} + T - 1$, $\hat{\mSigma} = \left( \sum_{t=2}^{T}\vv_{t} \vv_{t}^{\prime} + \mSigma_{0}^{-1} \right)^{-1}$, $\vv_{t} = \vbeta_{t}^{(m)} - \vbeta_{t-1}^{(m)}$.

These parameters are easily sampled by Gibbs sampler \citep[see][]{GS90}.

\section{Empirical Results and Distributional Dynamics}\label{sec:emp} 
\subsection{Dynamics of the Estimated Income Distribution}
Before examining empirics, we will explain the data, which is utilized in this study.
For an income data, we use a same dataset with \citet{NKO12,NK15}.
It is from the family income and expenditure survey prepared by the Statistics Bureau, Ministry of Internal Affairs and Communications.
The data are yearly data from 1969 to 2007 based on calender years, including the data of two types of households: workers' and two-or-more person households.
Other than the two types, there are the data called all households, which include single-person household in addition to the two-or-more person households.
However, it is available only from 1995, although this type of data is suitable to this analysis.
Therefore, we use the data of two-or-more person's households, which consist of workers households and non-workers' household including agricultural, forestry, and fisheries households, because we need longer data for time series analysis.
This data is offered by the quintile form.

\begin{figure}[tbp]
	\centering
	\includegraphics[width=\linewidth]{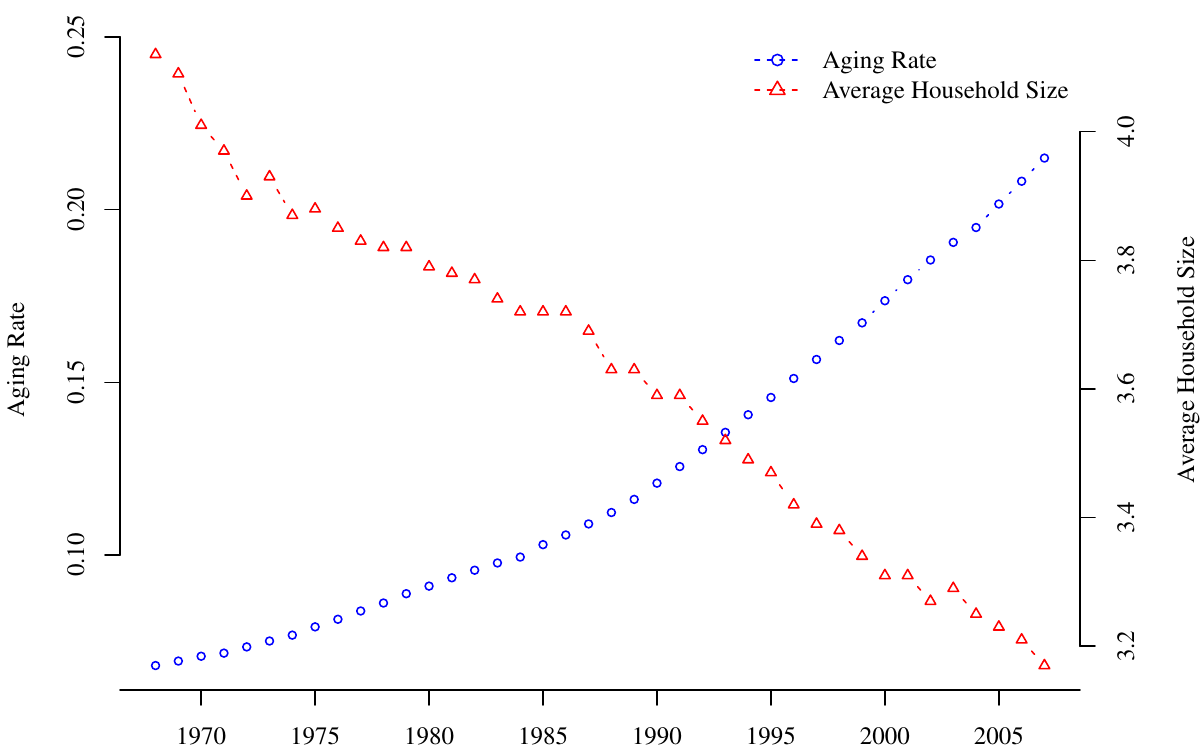}
	\caption{The aging rate and average household size}
	\label{fig:covariates}
\end{figure}

\begin{center}[INCLUDE \autoref{fig:covariates} HERE]\end{center}

\begin{figure}[tbp]
	\centering
	\includegraphics[width=\linewidth]{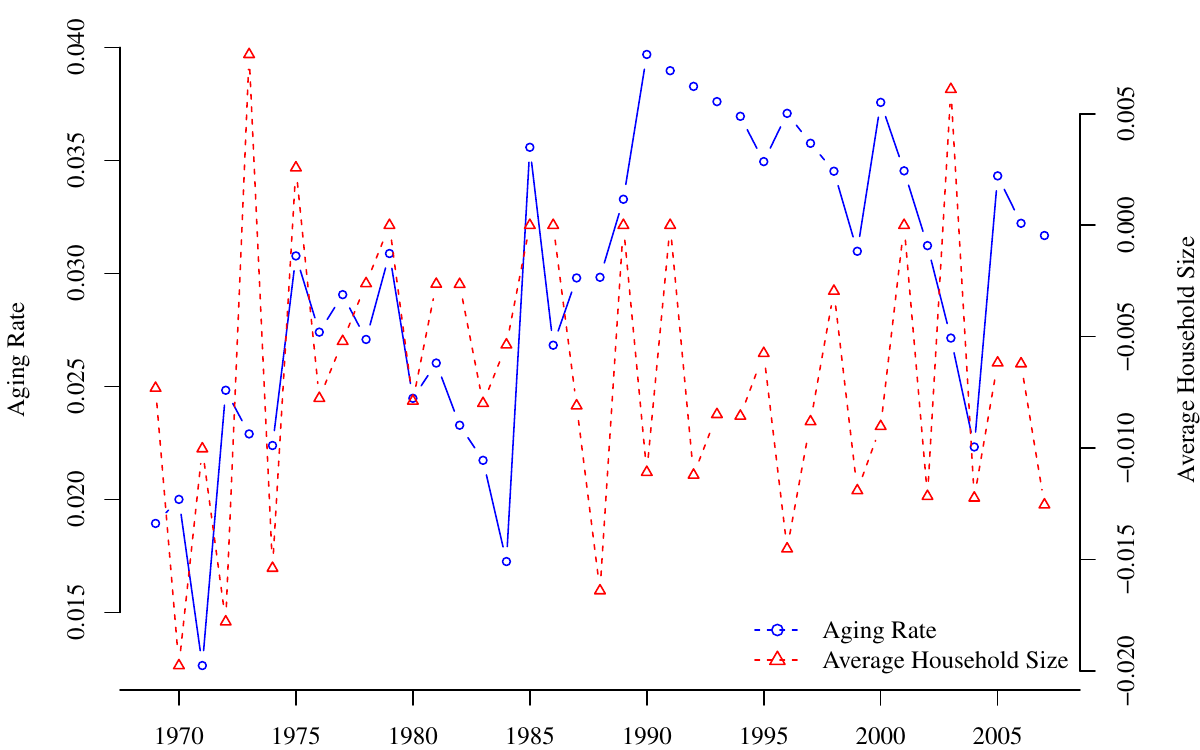}
	\caption{The log-difference of the aging rate and average household size}
	\label{fig:covariates-diff}
\end{figure}

\begin{center}[INCLUDE \autoref{fig:covariates-diff} HERE]\end{center}

As explanatory variables, we use aging rate and average household size from Population Estimates prepared by the Statistics Bureau, Internal Affairs and Communications and comprehensive survey of living conditions prepared by Ministry of Health, Labour and Welfare, respectively.
We use two demographic covariates, and therefore, $d=2$ in this empirical analysis.
Aging rate is defined by the share of over 65 years person in total population and average household size is an average of household members in two-or-more person households.
Figure \ref{fig:covariates} shows the trend of these variables.
From the figure, we can confirm the aging of the population and the decrease in the number of household members.
Since the aging rate and the average household size exhibit upward and downward trends, respectively, the variable for period $t$ is defined as the first difference of the logarithm of each variable.
Figure \ref{fig:covariates-diff} shows the log-difference of these variables.
In this analysis, we examine these effects on income inequality.

Given a dataset, we run the MCMC algorithm using $200,000$ iterations and discarding the first $50,000$ iterations as a burn-in period.
With the remaining $150,000$ samples, we store $15,000$ thinned samples every 10th draw after the initial burn-in period.
Moreover, we set the hyper-parameters to $\vbeta_{0} = \vzero_{4d}$, $\mDelta_{0} = 100 \times \mI_{4d}$, $\vmu_{0} = \vzero_{4}$, $\mPhi_{0} = 100 \times \mI_{4}$, $n_{0} = 5$, $\mOmega_{0} = 1000 \times \mI_{4}$, $m_{0} = 4d + 1$ and $\mSigma_{0} = 1000 \times \mI_{4d}$, where $\vzero_{n} =(\underbrace{0,0,\ldots,0}_{n})^{\prime}$ and $\mI_{n}$ is an $n \times n$ unit matrix.
All the results reported here were generated using Ox version 9.30 (macOS\_64/Parallel) \citep[see][]{D13} and all the figures are drawn using R version 4.5.3 \citep[see][]{R26}.

To compare the proposed model given in \eqref{eqn-model-obs}--\eqref{eqn-model-state2} with the existing model, we also estimate the parameters of the model specified in \eqref{eqn-model-obs}, whose likelihood function is provided in \eqref{eqn-likelihood-obs}.
Hereafter, we refer to this model as the independent model.
Unlike our proposed model, we estimate $\exp(\vh_{t}) = \vtheta_{t} = (a_{t}, b_{t}, p_{t}, q_{t})^{\prime}$ instead of $\vh_{t}$.
Accordingly, we assume the following prior distribution:
\begin{align*}
	a_{t}, b_{t}, p_{t}, q_{t} \sim \G(\nu_{0}, \lambda_{0}),\quad \text{for}\ t=1,2,\ldots, T,
\end{align*}
where $\G$ denotes the gamma distribution, and the hyperparameters are set to $\nu_{0} = \lambda_{0} = 1$.
As in our proposed model, we run the MCMC algorithm of \citet{KN19} for $200,000$ iterations, discarding the first $50,000$ iterations as a burn-in period.
With the remaining $150,000$ samples, we store $15,000$ thinned samples every 10th draw after the initial burn-in period.

\begin{landscape}
	\begin{figure}[p]
		\centering
		\includegraphics[width=.45\linewidth]{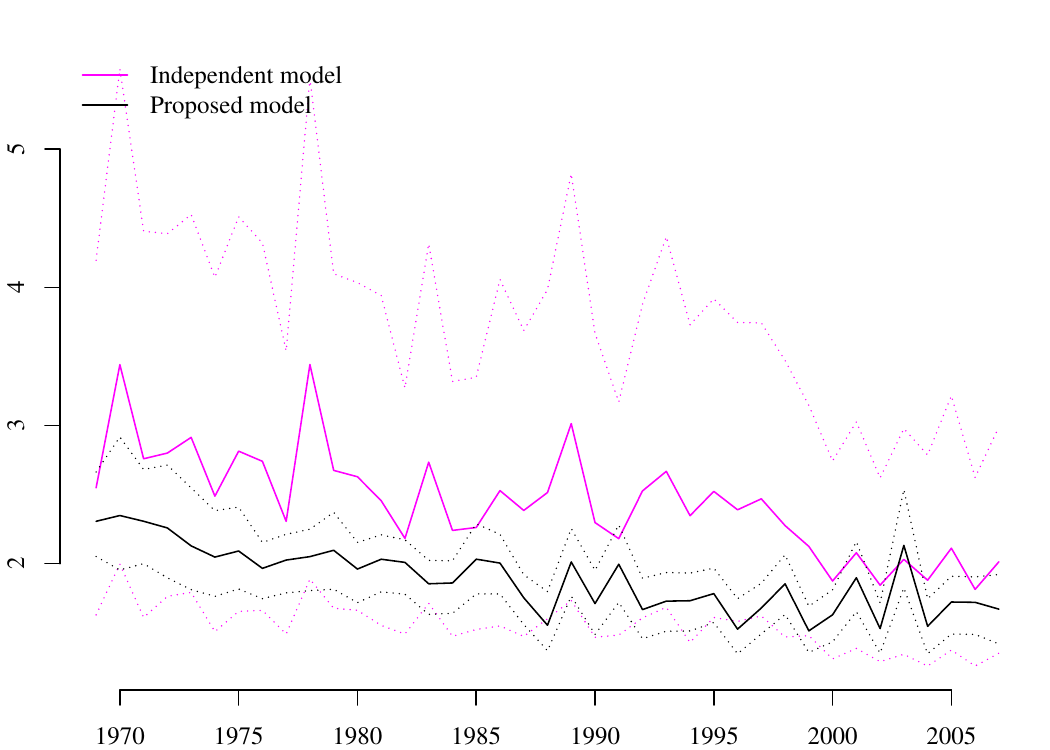}
		\includegraphics[width=.45\linewidth]{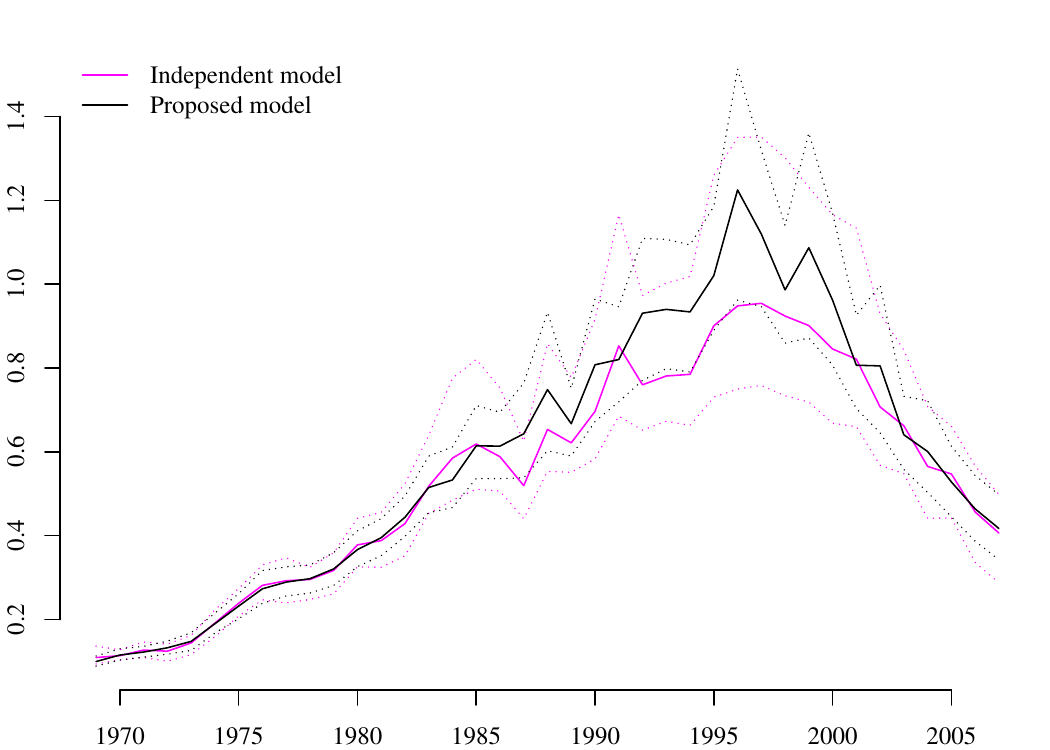}
		\includegraphics[width=.45\linewidth]{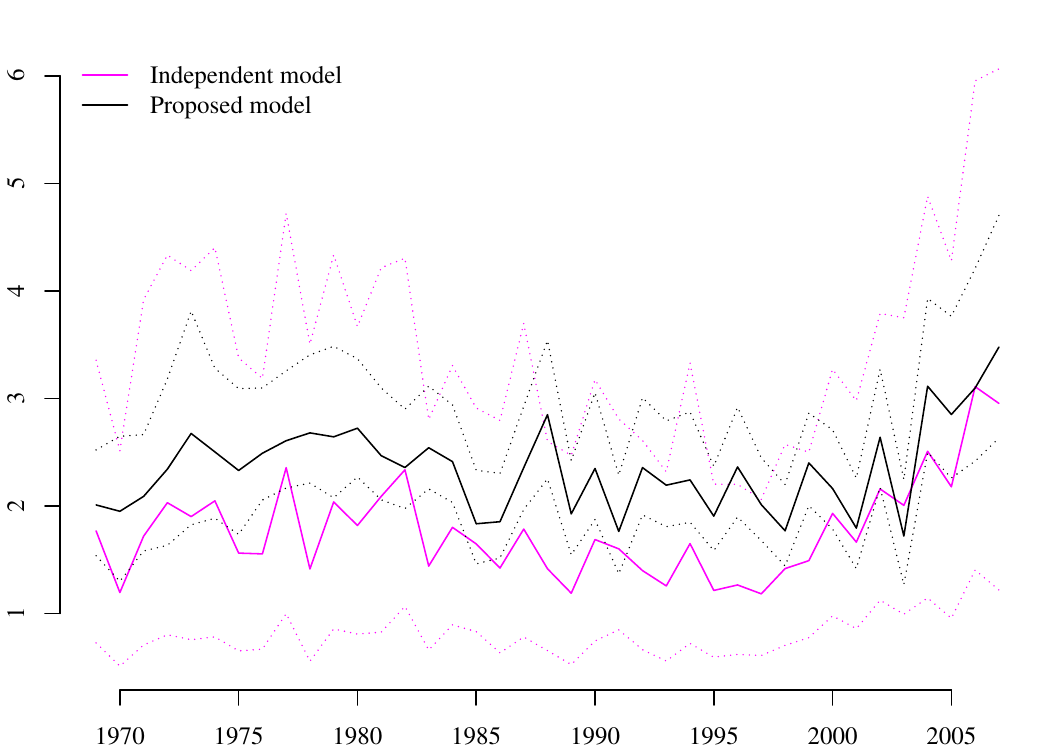}
		\includegraphics[width=.45\linewidth]{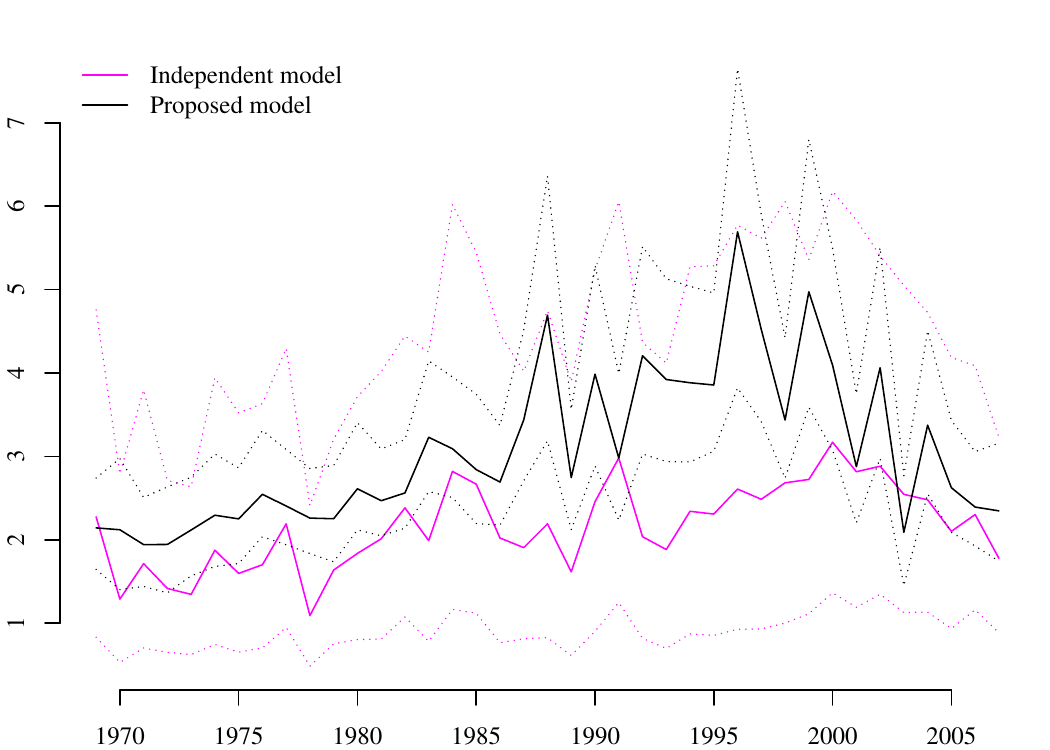}
		\caption{The posterior means and $95$\% credible intervals of the GB2 parameters (top-left: $a$, top-right: $b$, bottom-left: $p$, bottom-right: $q$)}
		\label{fig:ts}
	\end{figure}
\end{landscape}

\begin{center}[INCLUDE \autoref{fig:ts} HERE]\end{center}

Figure \ref{fig:ts} shows the posterior means with $95$\% credible intervals for the parameters of the GB2 distribution from both models.
With respect to parameter $a$, although the $2.5$th percentiles of the $95$\% credible intervals are nearly identical under both models, the $97.5$th percentiles are larger under the independent model than under the proposed model, suggesting that the posterior means are estimated to be larger under the independent model.
With respect to parameter $b$, although the $97.5$th percentiles of the $95$\% credible intervals are estimated similarly under both models, the $2.5$th percentiles under the independent model are smaller than those under the proposed model during the period from approximately 1985 to 2000.
Consequently, the posterior means under the independent model for this period are estimated to be smaller than those under the proposed model.
With respect to parameters $p$ and $q$, although the $97.5$th percentiles of the $95$\% credible intervals are estimated similarly under both models, the $2.5$th percentiles under the independent model are smaller than those under the proposed model.
Accordingly, the posterior means under the independent model are estimated to be slightly smaller overall than those under the proposed model.
Consequently, by employing a dynamic model that incorporates information from other periods in the parameter estimation, the credible intervals for all parameters become narrower, in particular through the narrowing of the credible interval for parameter $a$.
The differences in the posterior means between the two models are attributable to the reduction in uncertainty.
When viewed in terms of the posterior means, and taking the skewness of the posterior distribution into account, these differences may be regarded as an improvement in estimation accuracy.
Thus, as in \citet{KYKKS22,HHIS24}, we believe that the objective of improving estimation accuracy has been achieved by introducing the novel dynamic model.
Therefore, we focus on the changes in each parameter from our proposed model.

For parameter $a$, although it does not exhibit a smooth movement and instead shows short-term fluctuations, when viewed overall it appears to have a negative trend.
This may be interpreted as indicating a movement toward increasing inequality while thickening both the left and right tails of the distribution.
However, it should also be noted that the magnitude of this trend is small.
For parameter $b$, we can confirm that a positive trend can be observed until 1995 and it turns to a negative trend.
It should be mentioned that as it does not affect on income inequality, it is out of our concern.
However, it can be seen that this trend continued until after the collapse of the bubble economy.
With respect to parameter $p$, although there are short-term fluctuations, it appears to remain relatively stable at around 2 to 3 until around 2000.
Thereafter, it increases sharply.
In other words, after around 2000, the growth of the low-income population may be interpreted as being accompanied by greater equality within the low-income segment.
Considering the downward trend in parameter $a$, the extent to which equality among the low-income segment progressed should be discussed with caution.
However, focusing solely on $p$, an equalizing tendency after around 2000 can be observed.
Finally, focusing on parameter $q$, although short-term fluctuations are again observed, there appears to be a slight positive trend until mid-1990s, followed thereafter by a negative trend.
In other words, with attention to the high-income segment, inequality appears to have declined up to mid-1990s as the size of the high-income segment decreased, whereas after mid-1990s inequality can be interpreted as having increased through an expansion of the high-income segment.
However, as in the interpretation of parameter $p$, the movement of parameter $a$ must also be taken into account in order to accurately capture changes in inequality.
Nevertheless, since both $a$ and $q$ exhibit negative trends after mid-1990s, it may be concluded that inequality expanded through greater dispersion within the upper tail of the distribution.
Having identified the trends in the changes of the parameters, we next examine why these parameters changed.

\begin{figure}[tbp]
	\centering
	\includegraphics[width=.47\linewidth]{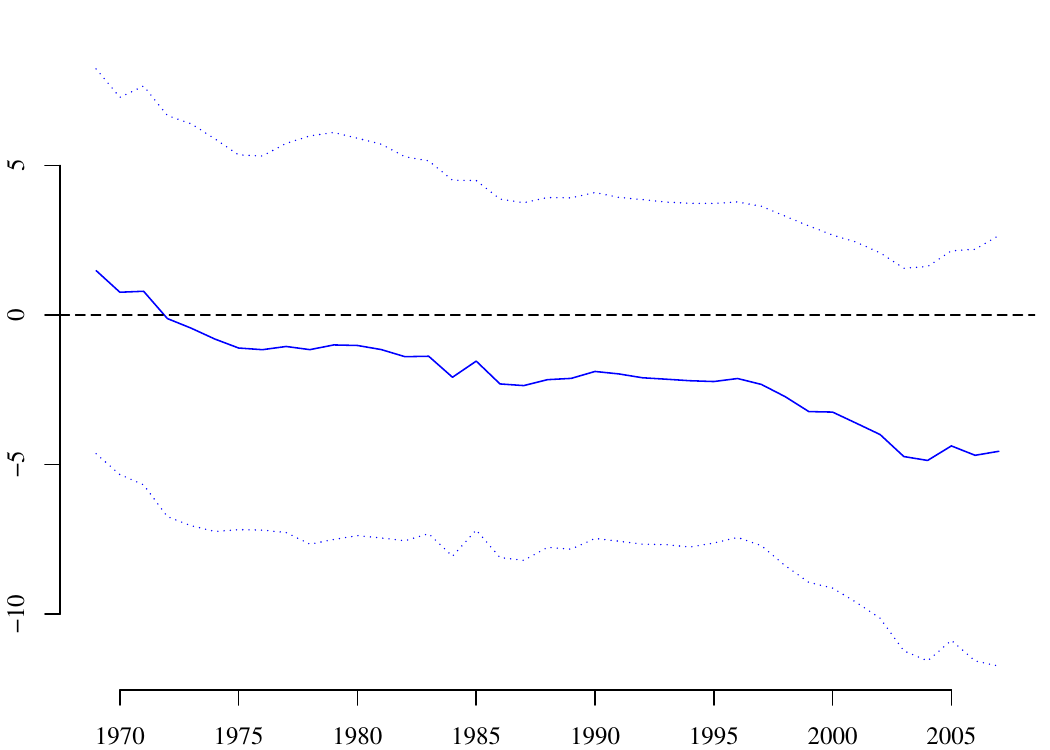}
	\includegraphics[width=.47\linewidth]{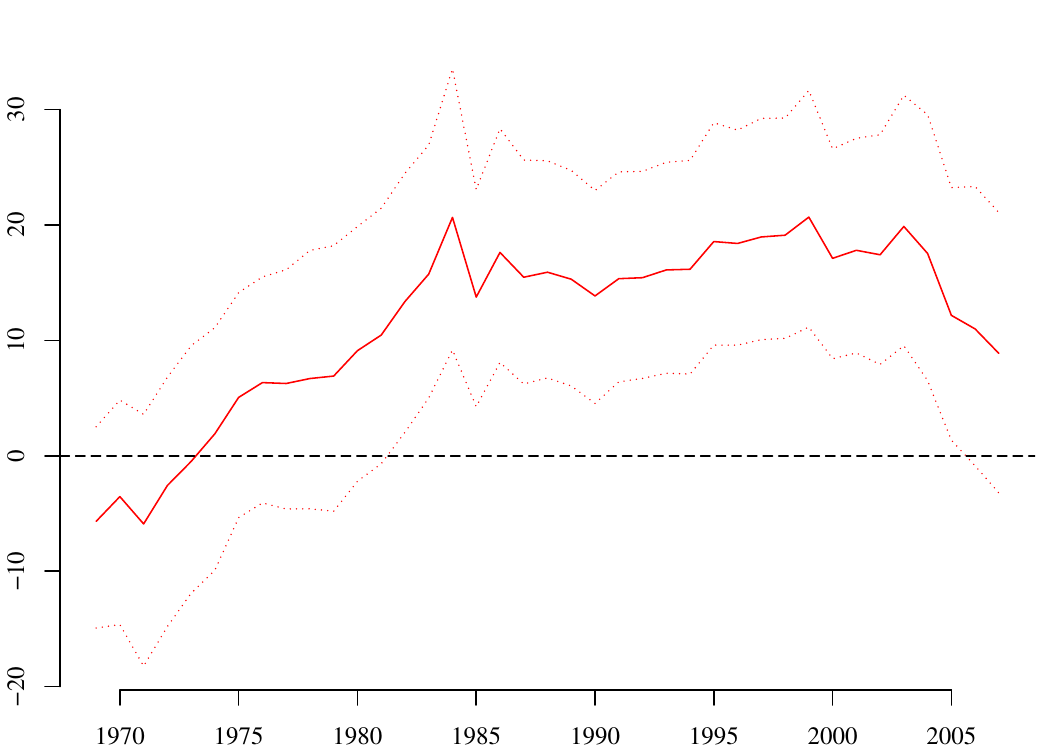}
	\includegraphics[width=.47\linewidth]{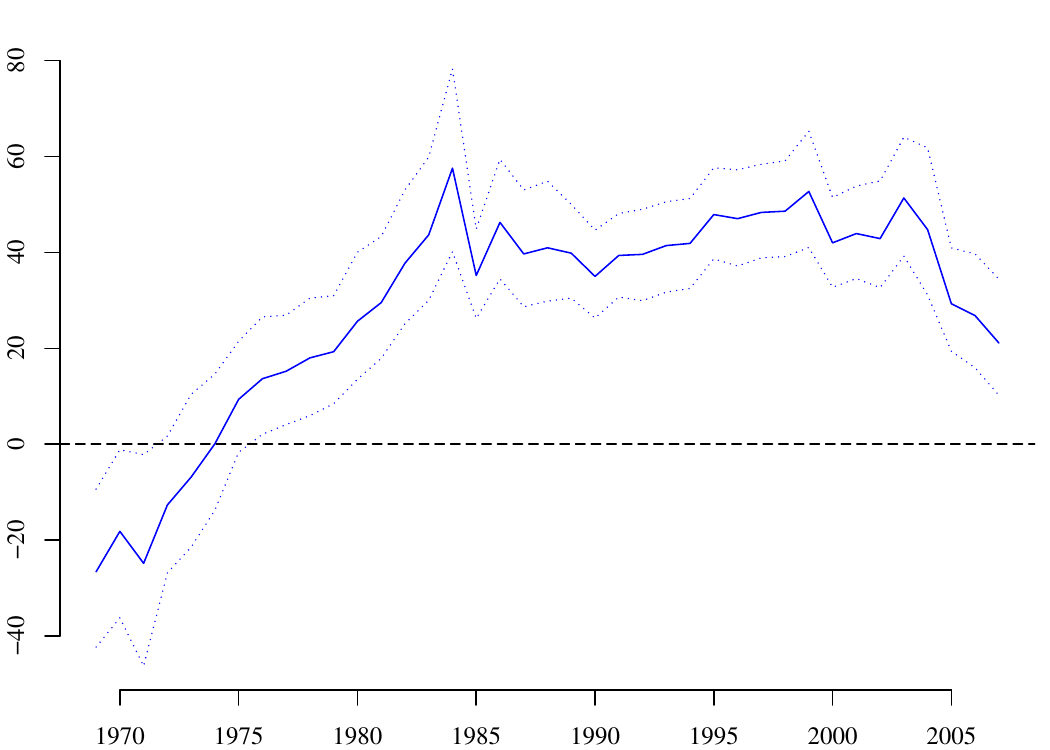}
	\includegraphics[width=.47\linewidth]{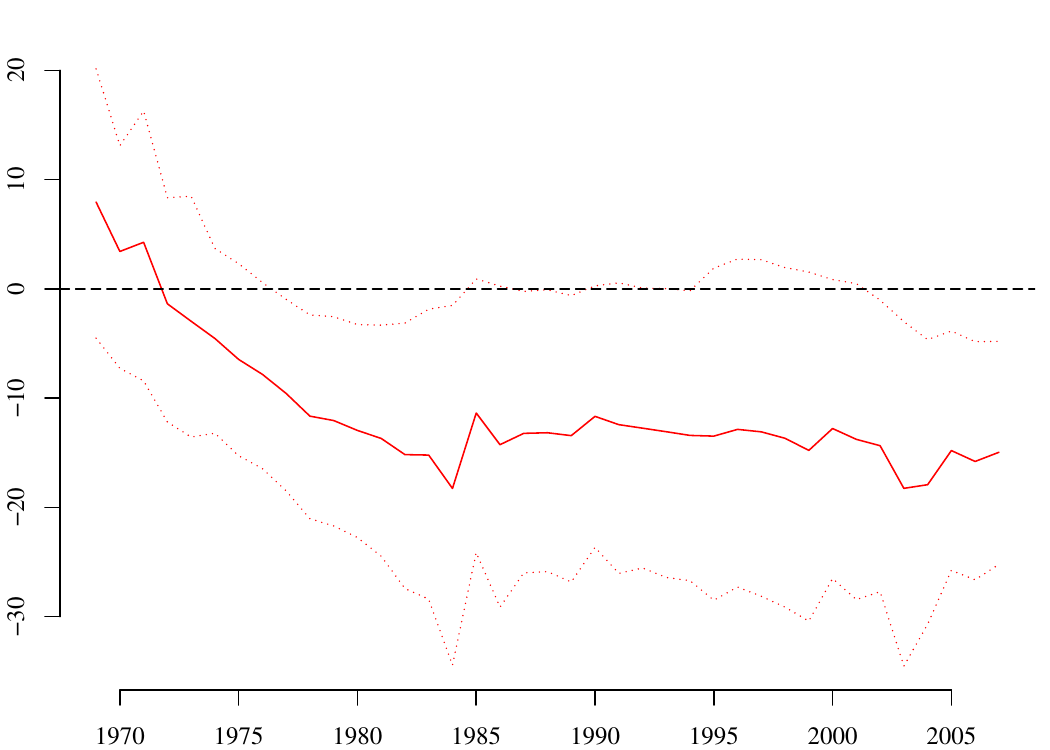}
	\includegraphics[width=.47\linewidth]{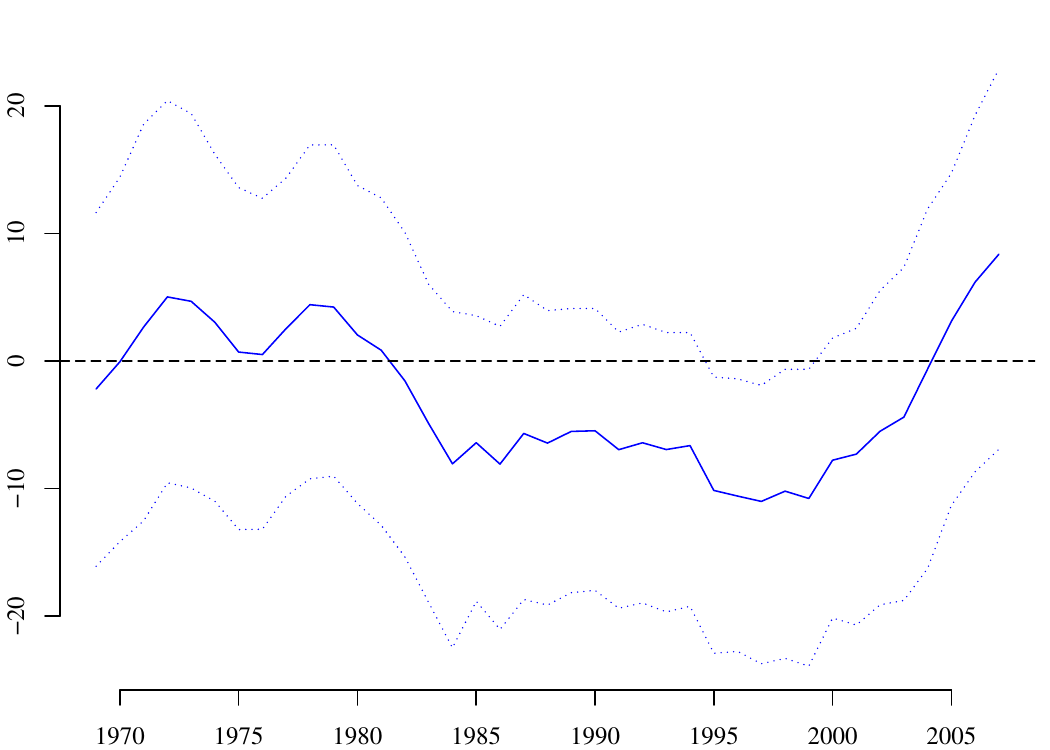}
	\includegraphics[width=.47\linewidth]{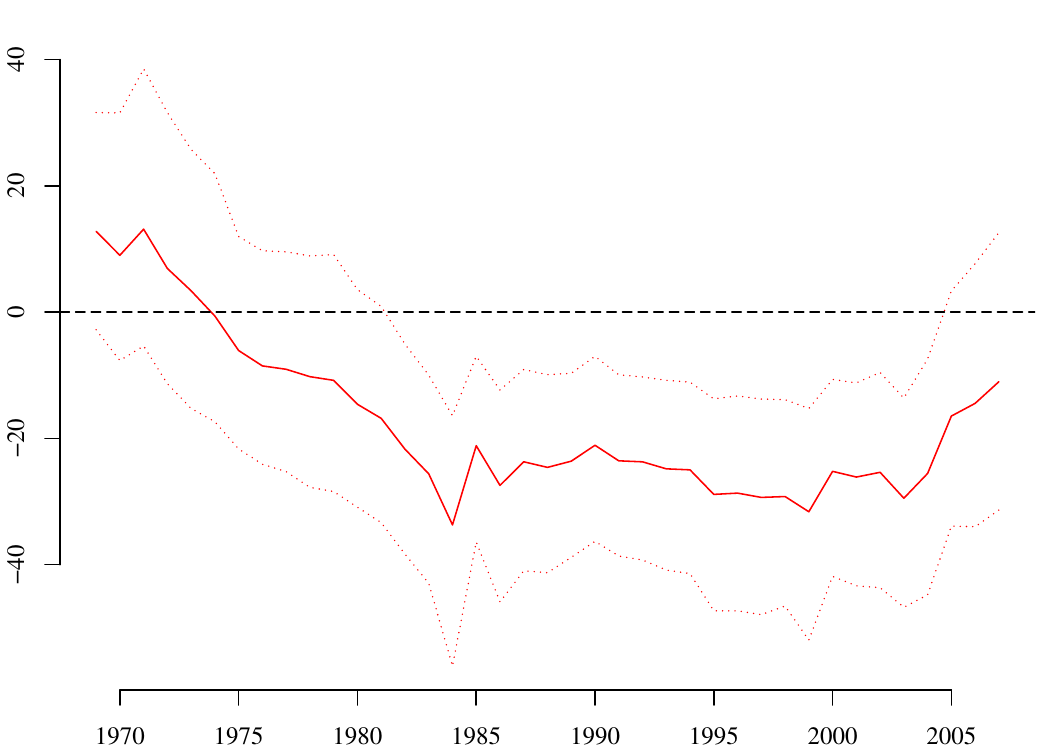}
	\includegraphics[width=.47\linewidth]{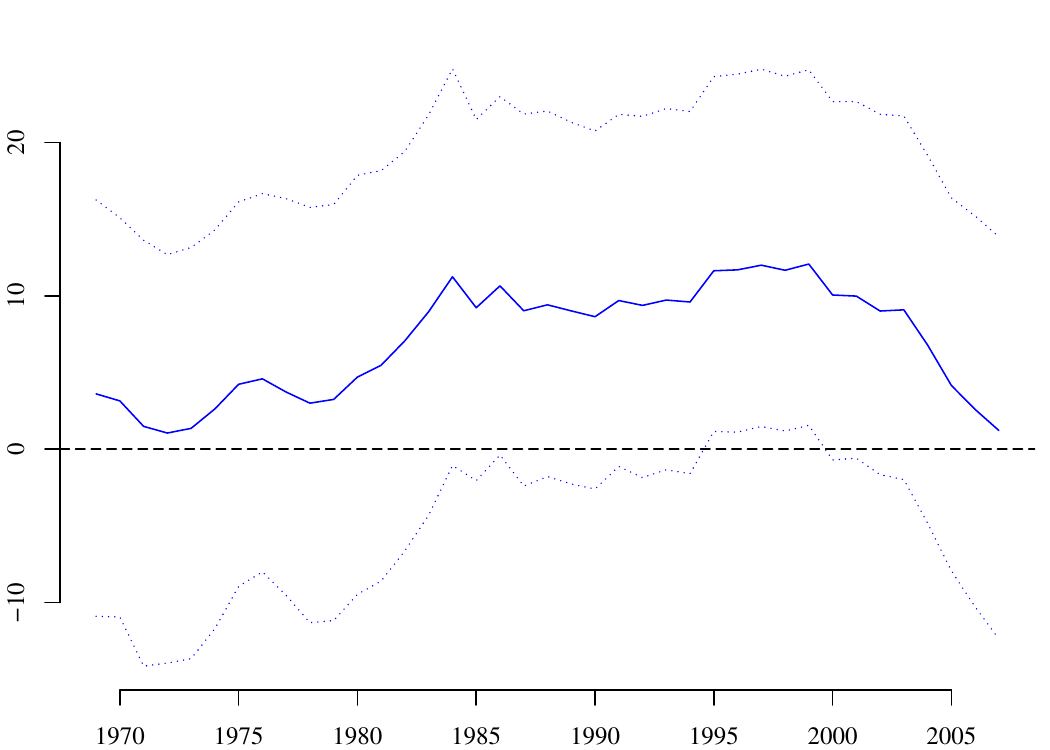}
	\includegraphics[width=.47\linewidth]{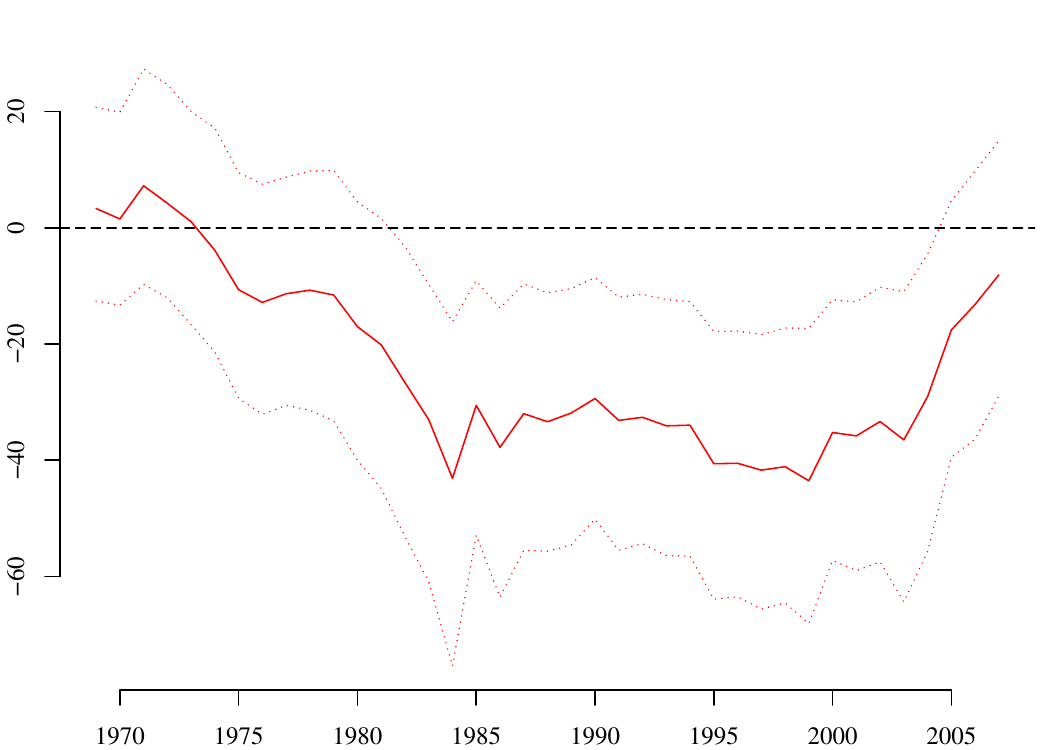}
	\caption{The posterior means and $95$\% credible intervals of the regression coefficients associated with the aging rate (left) and average household size (right) (From top to bottom, $a$, $b$, $p$, $q$)}
	\label{fig:beta}
\end{figure}

\begin{center}[INCLUDE \autoref{fig:beta} HERE]\end{center}

Figure \ref{fig:beta} shows the posterior means with $95$\% credible interval for $\vbeta_{t}$ for $t=1,2,\ldots,T$.
From the figure, viewed overall, for many parameters there are multiple periods in which the $95$\% credible intervals include zero, while there are also periods in which the $95$\% credible intervals do not include zero, and these patterns differ across parameters.
We therefore examine each parameter in detail.
Here, it should be noted that, as shown in Figure \ref{fig:covariates-diff}, the log difference of the aging rate, which is an explanatory variable, takes positive values, whereas the log difference of the average household size takes negative values for most of the period.

We first examine parameter $a$.
The coefficient for the aging rate appears to exhibit a negative trend; however, since the $95$\% credible interval includes zero throughout the entire period, its effect on parameter $a$ is considered to be limited.
Nevertheless, given that this explanatory variable takes positive values, it may still have contributed to the negative trend in parameter $a$.
Turning to the average household size, the corresponding coefficient has a $95$\% credible interval that includes zero through the early 1980s.
It then remains positive while following an upward trend in the mid-1980s, after which it stays positive and relatively stable through the early 2000s.
Thereafter, it declines, and by the mid-2000s the $95$\% credible interval comes to include zero.
Since the log difference of the average household size takes negative values, these results suggest that the decline in average household size contributed to the negative trend in parameter $a$, thereby serving as a factor behind the expansion of inequality.

We next turn to parameter $b$.
Since parameter $b$ lies outside the main focus of our analysis, we provide only a brief overview.
The coefficient for the aging rate exhibits an upward trend through the mid-1980s, after which it remains relatively stable.
From From the early 2000s onward, a slight downward trend can be observed.
By contrast, the coefficient for average household size follows a downward trend through the mid-1980s and thereafter remains relatively stable at negative values.
Taking into account the movements of these two explanatory variables, these results appear to explain the upward trend in parameter $b$ up to the mid-1990s.
However, they do not fully capture the subsequent downward trend in parameter $b$.

Third, we focus on parameter $p$.
Looking first at the coefficient for the aging rate, the posterior mean fluctuates around zero until 1982.
Thereafter, it remains negative through the early 2000s, and around the late 1990s the $95$\% credible interval does not include zero.
Since the coefficient then exhibits an upward trend from the late 1990s onward, changes in lower-tail inequality were observed in the mid-to-late 1990s.
Although the $95$\% credible interval includes zero from the late 1990s onward, the subsequent upward trend may indicate a contribution to greater equality within the lower tail of the distribution.
Turning to the coefficient for average household size, it follows a downward trend through the mid-1990s and then remains negative and relatively stable through the early 2000s.
Thereafter, it shows an upward trend.
Since the $95$\% credible interval does not include zero through the mid-1990s, it may be inferred that equality progressed during this period alongside changes in the lower tail of the distribution.

At the end of the parameter explanation, we turn to parameter $q$.
Interestingly, the coefficient of the aging rate for parameter $q$ moves in the opposite direction to that for parameter $p$.
The posterior mean fluctuates around zero through the late 1970s.
Thereafter, it remains positive through the late 1990s, and during this period the $95$\% credible interval does not include zero.
Since the coefficient then exhibits a downward trend from the late 1990s onward, these results suggest that greater inequality emerged within the upper tail of the distribution in the mid-to-late 1990s.
Although the $95$\% credible interval includes zero from the late 1990s onward, the subsequent downward trend may indicate a contribution to greater inequality within the upper tail of the distribution through a compression of that part of the distribution.
Turning to the coefficient for average household size, as in the case of parameter $p$, it follows a downward trend through the mid-1990s and then remains negative and relatively stable through the early 2000s.
Thereafter, it exhibits an upward trend.
Since the $95$\% credible interval does not include zero through the mid-1990s, it may be inferred that equality progressed during this period while the upper tail of the distribution became thinner.

To sum up, by analyzing the dynamic model incorporating this regression structure, we were able to confirm from the model that population aging and the decline in average household size have affected the worsening inequality of income distribution in Japan.
In particular, one contribution of this analysis is the finding that the effects of these variables on the parameters differ across parameters and over time.
Overall, the model suggests that the current state of income inequality in Japan is characterized by an expansion of inequality in which the lower tail of the distribution became thicker while the upper tail became thinner, driven by population aging and the decline in average household size.

\begin{figure}[tbp]
	\centering
	\includegraphics[width=\linewidth]{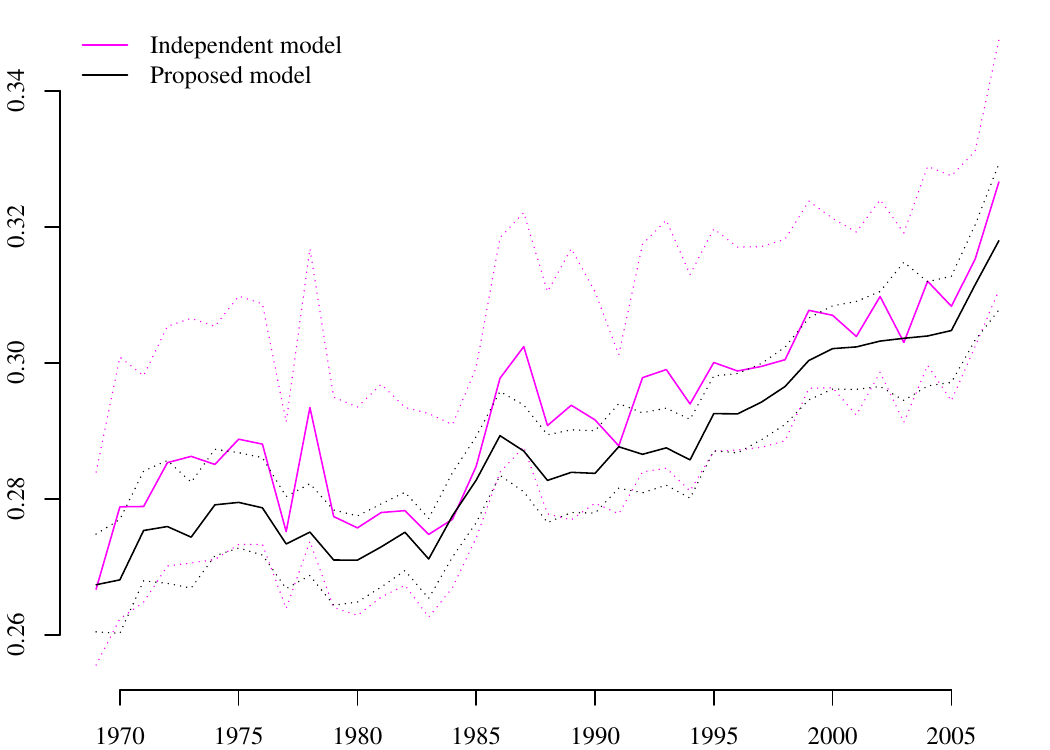}
	\caption{The posterior means and $95$\% credible intervals of the Gini coefficients}
	\label{fig:Gini}
\end{figure}

\begin{center}[INCLUDE \autoref{fig:Gini} HERE]\end{center}

Figure \ref{fig:Gini} shows the posterior means with $95$\% credible intervals for the Gini coefficients.
Focusing on the $95$\% credible intervals of the Gini coefficient, the $2.5$th percentiles under the independent model exhibit a trend very similar to those under the proposed model, whereas the $97.5$th percentiles under the independent model tend to be larger than those under the proposed model.
As a result, the posterior means of the Gini coefficient under the proposed model are estimated to be smaller than those under the independent model.
However, since the Gini coefficients under both models display broadly similar trends, the proposed model appears to achieve more precise estimation by reducing uncertainty through the use of information from other periods, as was also observed in the parameter estimation as in \citet{KYKKS22,HHIS24}.
Furthermore, it is noteworthy that the estimated Gini coefficient under the proposed model follows a smoother trajectory than that under the independent model.

Moreover, this figure is close to that of \citet{NKO12}, which assumes the lognormal distribution as the hypothetical income distribution with the same dataset, although the Gini coefficients from our model are slightly smaller than those from \citet{NKO12}.
It suggests that the assumption of the lognormal distribution is enough to examine the trend of income inequality for this data.
However, by assuming the GB2 distribution, we can also examine the cause of the changes in the income inequality in more detail although \citet{NKO12} fail to find the cause of the income inequality.
Therefore, this proposed model provides richer distributional insights than \citet{NKO12} in this sense.

\begin{figure}[tbp]
	\centering
	\includegraphics[width=\linewidth]{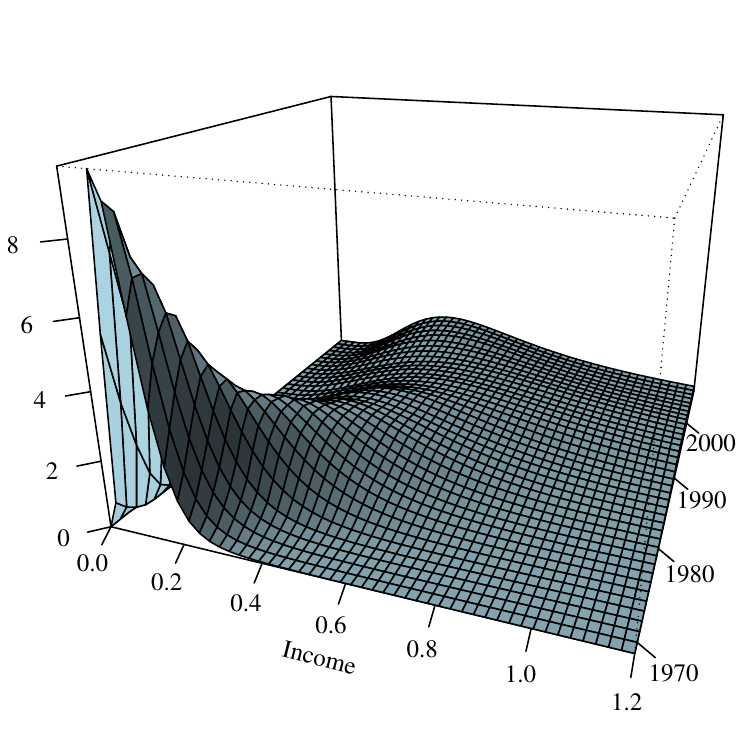}
	\caption{Income distribution and its evolution over time}
	\label{fig:ID}
\end{figure}

\begin{center}[INCLUDE \autoref{fig:ID} HERE]\end{center}

Finally, we examine the shape of the income distribution and its evolution over time in order to confirm the changes in the parameters and the trend in the Gini coefficient that were not fully apparent from those measures alone.
The income distribution for each period is drawn using the posterior means.
Figure \ref{fig:ID} presents the income distribution and its transition over time.
From this figure, it can be seen, for example, that the mode shifted toward the higher-income side up to around 1995 and then moved slightly back toward the lower-income side thereafter.
It can also be confirmed that the distribution shifted slightly toward the lower-income range.
Thus, the patterns inferred from the movements of the parameters and the Gini coefficient can also be visually verified from this figure.

\subsection{Counterfactual  Analysis of Demographic Effects}

While the empirical analysis identifies how demographic variables affect the latent GB2 parameters, it does not directly reveal how these effects translate into changes in the income distribution or summary measures of inequality.
To address this issue, we construct model-based counterfactual income distributions by removing the contribution of a selected demographic covariate from the estimated latent-state equation while preserving all remaining components of the estimated model.
Unlike approaches based on reweighting or distributional decomposition, our counterfactuals are generated within the estimated Bayesian state-space model and therefore retain the estimated temporal dependence of the latent distributional parameters.

Let $\vx_{(\ell),t}$ denote the vector $\vx_{t}$ with its $\ell$th element replaced by its counterfactual value (zero in the present application), and define $\mZ_{(\ell),t}=\mI_{4}\otimes\vx_{(\ell),t}^{\prime}$.
Using the $m$th posterior draw of the latent states and regression coefficients, $\mH^{(m)}$ and $\mB^{(m)}$, we construct the counterfactual latent state as
\begin{align*}
    \vh_{(\ell),t}^{(m)}
    =
    \vh_{t}^{(m)}
    -
    \mZ_t\vbeta_t^{(m)}
    +
    \mZ_{(\ell),t}\vbeta_t^{(m)},
    \qquad
    t=1,\ldots,T.
\end{align*}
This operation modifies only the contribution of the selected demographic covariate.
The estimated regression coefficients, intercept, latent-state realizations, residual shocks, and all remaining model components are preserved for each posterior draw.
Consequently, the model is not re-estimated under the counterfactual scenario; instead, the counterfactual distributions are obtained by altering only the structural contribution of the selected demographic variable within the estimated state-space system.

For each posterior draw, the counterfactual GB2 parameters are recovered as
$\exp\{\mH_{(\ell)}^{(m)}\}$,
from which the corresponding income distributions, Lorenz curves, and Gini coefficients are computed.
Because each GB2 parameter governs a different feature of the income distribution---overall concentration ($a$), scale ($b$), and the upper and lower tails ($p$ and $q$)---the proposed approach allows us to examine not only changes in aggregate inequality but also how demographic factors reshape different parts of the income distribution.

Our counterfactual framework differs fundamentally from reweighting-based methods (\citealp{DFL96}) and quantile-based counterfactual approaches (\citealp{CFM13}).
Rather than modifying the empirical distribution of the observed data, we alter the structural determinants entering the latent-state equation while preserving the estimated dynamic evolution of the remaining GB2 parameters.
The proposed framework therefore complements the broader literature on distributional decomposition (\citealp{FLF11,J95}) by providing a unified Bayesian approach to model-based counterfactual analysis for dynamic income distributions estimated from grouped data.

\begin{landscape}
	\begin{figure}[p]
		\centering
		\includegraphics[width=.45\linewidth]{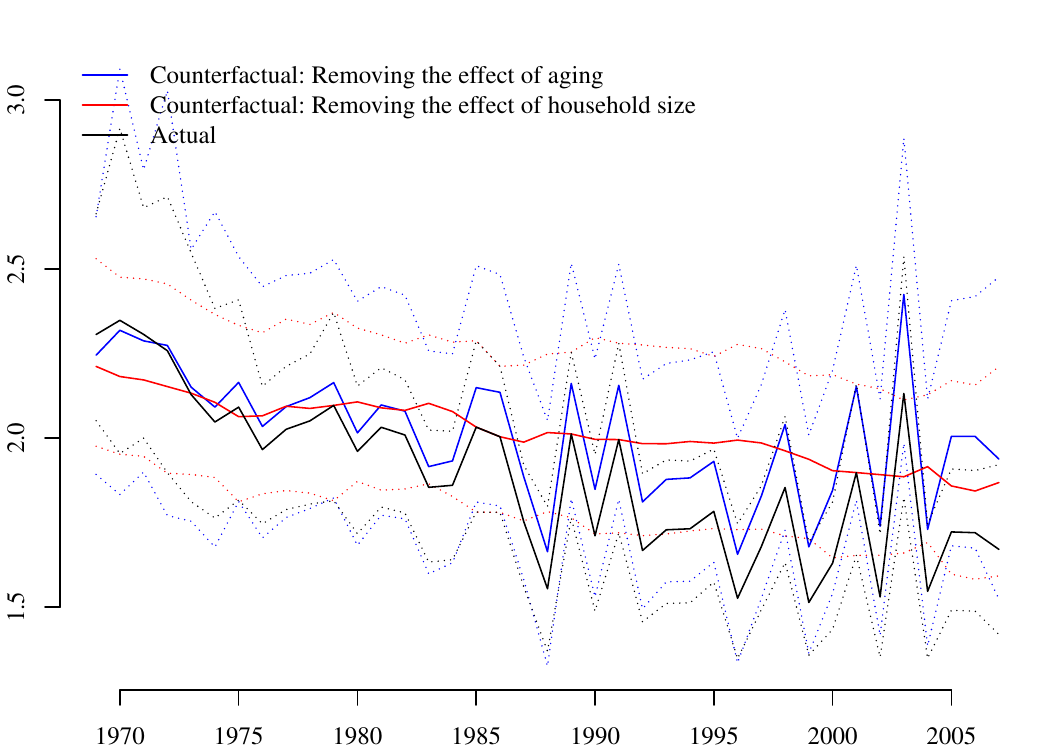}
		\includegraphics[width=.45\linewidth]{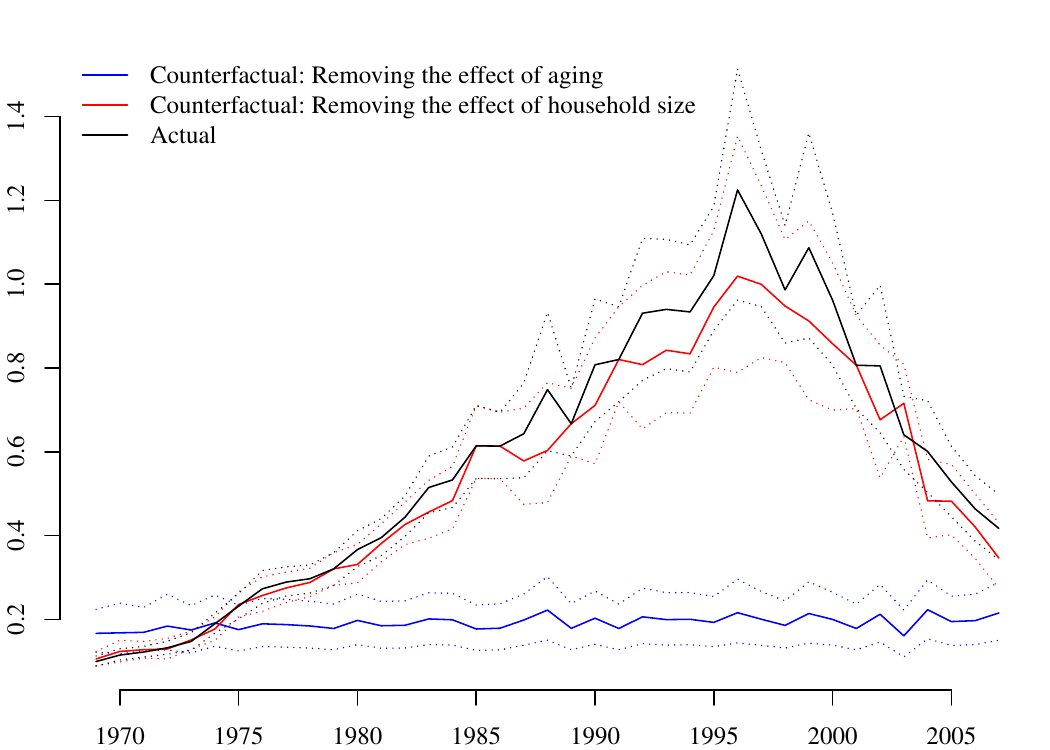}
		\includegraphics[width=.45\linewidth]{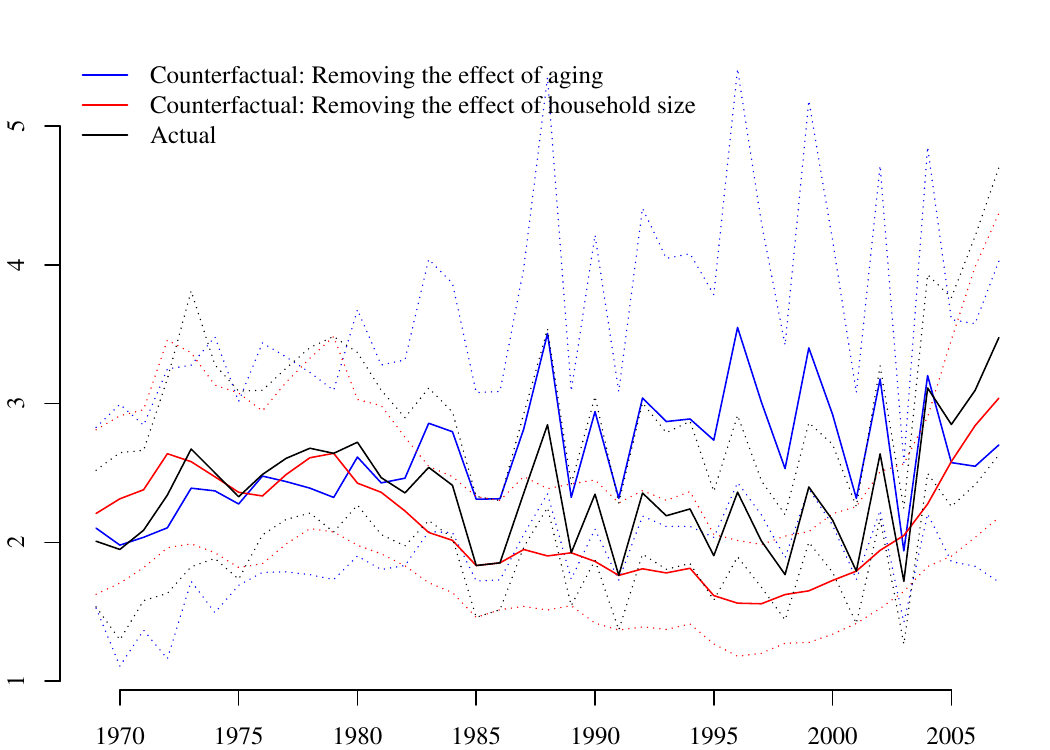}
		\includegraphics[width=.45\linewidth]{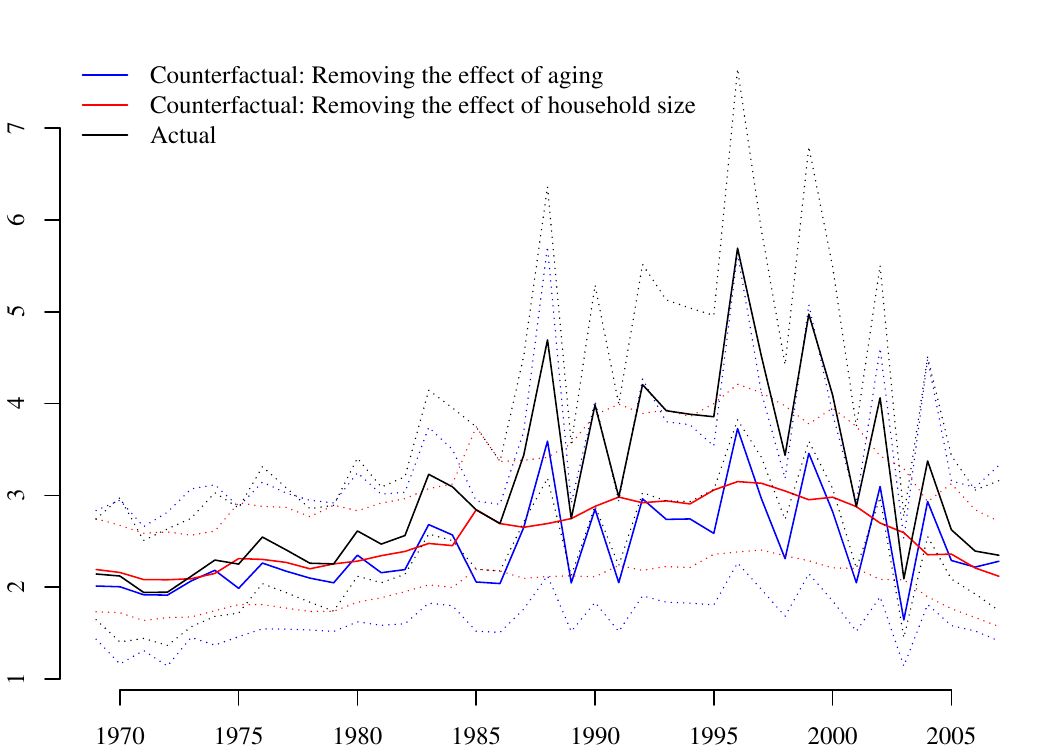}
		\caption{The counterfactual posterior means and $95$\% credible intervals of the GB2 parameters (top-left: $a$, top-right: $b$, bottom-left: $p$, bottom-right: $q$)}
		\label{fig:ts-cf}
	\end{figure}
\end{landscape}

\begin{center}[INCLUDE \autoref{fig:ts-cf} HERE]\end{center}

Figure \ref{fig:ts-cf} presents the trends in the counterfactual and actual parameters.
Focusing first on parameter $a$, the trend in the counterfactual parameter excluding the effect of population aging is similar to that of the actual parameter, although the counterfactual values are generally slightly larger.
In contrast to the aging counterfactual, the counterfactual path without changes in household size exhibits substantially smoother dynamics for parameter $a$.
This result suggests that the decline in household size played an important role not only in lowering the level of a, thereby contributing to increasing inequality, but also in generating short- to medium-run fluctuations in the dispersion of the income distribution.
Since lower values of a correspond to thicker lower and upper tails in the GB2 distribution, the results imply that the decline in household size contributed to a broader dispersion of income away from the middle-income group.

Next, we examine the trend in parameter $b$.
The results indicate that, even under the assumption that household size remained constant, the trend in the counterfactual parameter does not differ substantially from that of the actual parameter.
In contrast, under the assumption that population aging did not progress, the parameter remains largely unchanged over time.
These findings suggest that the changes in parameter $b$ were primarily driven by population aging.

Looking at parameter $p$, under the counterfactual scenario in which population aging did not progress, the parameter follows a trajectory broadly similar to the observed one, although the values are slightly higher from the 1980s onward.
Since larger values of $p$ correspond to a thinner lower tail of the GB2 distribution, this result suggests that population aging contributed to a thickening of the lower tail of the distribution.
By contrast, under the counterfactual scenario in which the decline in household size did not occur, the trajectory of $p$ remains close to the observed level but evolves more smoothly over time.
This finding implies that the decline in household size mainly contributed to short- and medium-run fluctuations in the lower-tail dynamics rather than to persistent shifts in the level of $p$.

Finally, we examine the trend in parameter $q$.
Under the assumption that population aging did not occur, the trend remains similar to the actual one, although the values are generally smaller.
In contrast, under the assumption that household size did not decline, the trend becomes considerably smoother.
These results suggest that population aging primarily affected the level of the upper tail of the income distribution, tending to reduce inequality within the upper tail of the distribution, whereas the decline in household size mainly contributed to short- and medium-term fluctuations in upper-tail dynamics.

The counterfactual results for parameters $a$, $p$, and $q$ suggest that population aging and the decline in household size affected different aspects of the income distribution in Japan.
First, the decline in household size appears to have played an important role in generating short- and medium-run fluctuations in the distributional dynamics, as the counterfactual paths of $a$, $p$, and $q$ become substantially smoother when changes in household size are removed.
Since lower values of a correspond to thicker lower and upper tails of the GB2 distribution, these results imply that the decline in household size contributed to a broader dispersion of income away from the middle-income group.
In addition, the counterfactual results for $p$ indicate that population aging contributed to a thickening of the lower tail of the distribution, whereas the results for $q$ suggest that aging simultaneously reduced inequality within the upper tail of the distribution by compressing the upper tail of the distribution.
Overall, the results imply that demographic changes in Japan shifted the income distribution toward a thicker lower tail while also altering the dynamics of both the lower and upper tails of the distribution.

\begin{figure}[tbp]
	\centering
	\includegraphics[width=\linewidth]{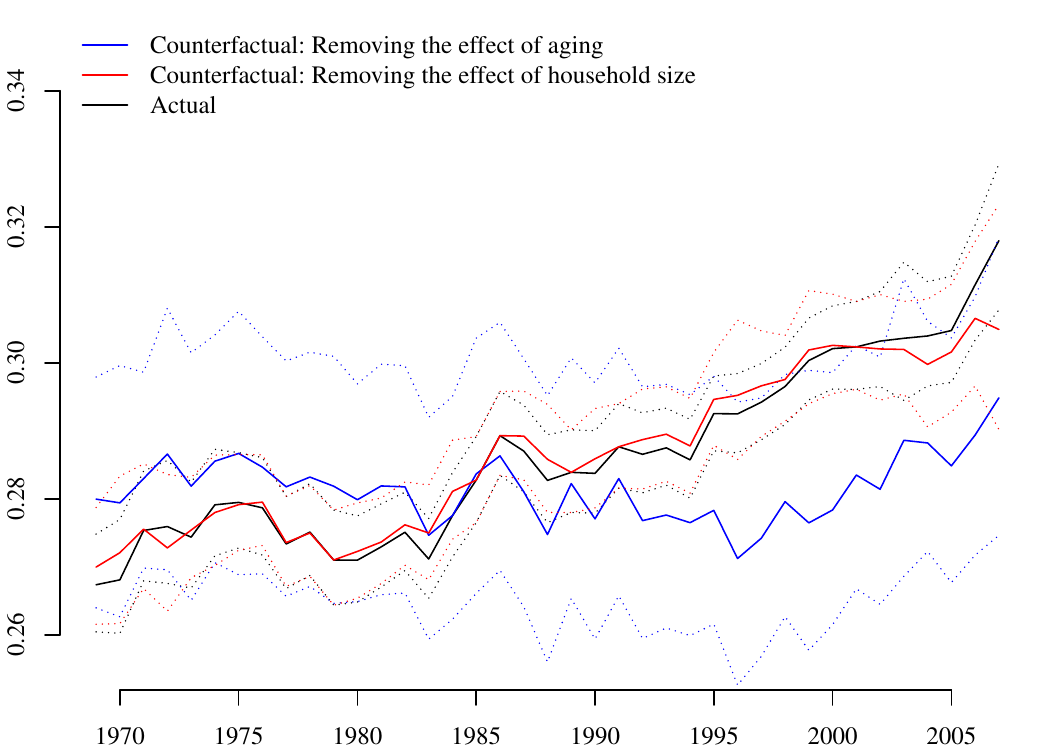}
	\caption{The actual and counterfactual posterior means and $95$\% credible intervals of the Gini coefficients}
	\label{fig:Gini-cf}
\end{figure}

\begin{center}[INCLUDE \autoref{fig:Gini-cf} HERE]\end{center}

Figure \ref{fig:Gini-cf} presents the counterfactual analysis of the Gini coefficient in order to examine how the changes in the GB2 parameters affected income inequality.
The figure shows that, even under the counterfactual scenario in which the decline in household size did not occur, the Gini coefficient follows a trajectory broadly similar to the observed one.
By contrast, under the counterfactual scenario without population aging, the Gini coefficient is slightly larger than the observed value before around 1985, but becomes smaller thereafter and remains relatively stable over time.
These results suggest that the effect of population aging on income inequality changed over time, possibly reflecting different impacts on the lower and upper tails of the income distribution.
Although the posterior means differ across scenarios, the $95$\% credible intervals overlap substantially.

\begin{figure}[tbp]
	\centering
	\includegraphics[width=\linewidth]{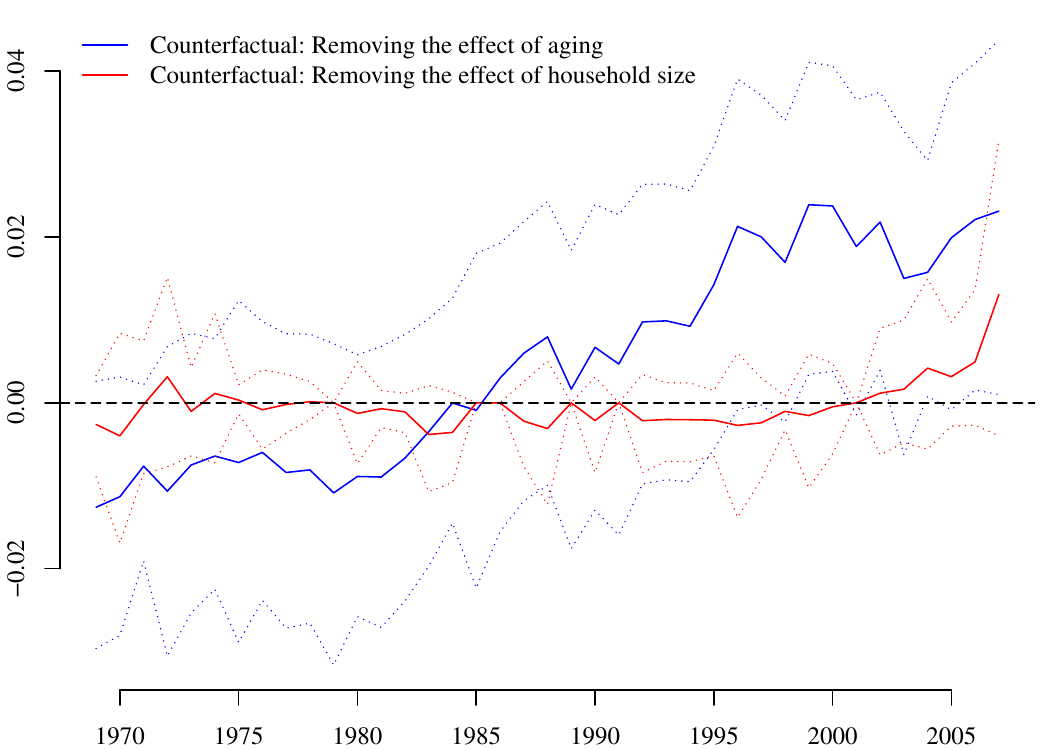}
	\caption{The posterior means and $95$\% credible intervals of the differences between the observed and counterfactual Gini coefficients}
	\label{fig:Gini-diff}
\end{figure}

\begin{center}[INCLUDE \autoref{fig:Gini-diff} HERE]\end{center}

To quantify these differences, Figure \ref{fig:Gini-diff} presents the posterior distributions of the differences between the observed Gini coefficient and the counterfactual Gini coefficients.
The results indicate that the effect of the decline in household size on the Gini coefficient is close to zero throughout the sample period, with the corresponding $95$\% credible intervals generally including zero.
By contrast, the effect of population aging becomes increasingly pronounced after around 2000, and the $95$\% credible intervals lie predominantly away from zero, indicating that aging played a meaningful role in shaping income inequality during this period.

These findings highlight the importance of modeling the entire income distribution rather than relying solely on summary inequality measures.
Although the decline in household size has only a limited effect on the Gini coefficient, it substantially alters the dynamics of the GB2 parameters, leading to noticeable changes in the shape of the income distribution, particularly in its lower and upper tails.
Conversely, population aging primarily induces persistent changes in the distributional shape that ultimately translate into long-run increases in aggregate inequality.
Taken together, these results demonstrate that demographic factors can reshape the income distribution through different distributional channels, even when their effects on conventional inequality measures appear similar.

To sum up, the counterfactual analyses suggest that population aging and the decline in household size affected different aspects of the income distribution in Japan.
The results for parameters $a$, $p$, and $q$ indicate that the decline in household size mainly contributed to short- and medium-run fluctuations in the dynamics of the income distribution, as the counterfactual parameter paths become substantially smoother when changes in household size are removed.
In contrast, population aging primarily affected the levels of the distributional parameters, contributing to a thickening of the lower tail while simultaneously compressing the upper tail of the distribution.
However, despite these effects on the shape and dynamics of the distribution, the counterfactual analysis of the Gini coefficient indicates that the decline in household size had only a limited effect on aggregate income inequality throughout the sample period.
By contrast, the effect of population aging on the Gini coefficient became increasingly pronounced after around 2000, with the corresponding $95$\% credible intervals excluding zero.
Overall, these results suggest that population aging was the main demographic factor driving the long-run increase in income inequality in Japan, whereas the decline in household size mainly affected the short-run distributional dynamics rather than the overall level of inequality.

\section{Conclusions}\label{sec:conclusion}

This paper proposed a Bayesian state-space framework for estimating dynamic income distributions from grouped income data and illustrated its usefulness through an application to demographic changes in Japan.
By combining a flexible GB2 specification with a dynamic state-space model, the proposed framework exploits temporal dependence in the evolution of income distributions, thereby improving estimation stability under grouped observations.
The framework further incorporates demographic covariates into the latent-state dynamics, enabling model-based counterfactual analyses of the factors driving changes in income distributions.
In this respect, the proposed approach extends the dynamic distributional models of \citet{KYKKS22,HHIS24} by integrating regression-based structural analysis within a unified Bayesian framework.

The empirical results revealed that demographic changes affected different aspects of the income distribution through different GB2 parameters.
In particular, population aging contributed to a thickening of the lower tail of the distribution while simultaneously compressing the upper tail, whereas the decline in household size mainly contributed to short- and medium-run fluctuations in distributional dynamics.
The counterfactual analyses further showed that, despite its influence on the evolution of the income distribution, the decline in household size had only a limited effect on the aggregate Gini coefficient.
By contrast, the contribution of population aging to the Gini coefficient became increasingly pronounced after around 2000, suggesting that demographic transition was an important driver of the long-run increase in income inequality in Japan.

These findings demonstrate that changes in demographic structure can reshape different parts of the income distribution in ways that are not fully captured by aggregate inequality measures alone.
Our results also complement the historical evidence documented by \citet{MS08}, who showed that postwar Japan experienced substantial changes in the concentration and composition of top incomes.
Their finding that the upper tail changed considerably despite relatively modest movements in summary inequality measures is consistent with our result that demographic factors affect different parts of the income distribution through distinct distributional channels.
More generally, the proposed framework provides a model-based approach for linking observed changes in grouped income distributions to their underlying structural determinants.
Because it requires only repeated grouped income data, the methodology is applicable even when individual-level microdata are unavailable and is therefore readily transferable to official income statistics in many other countries.
Taken together, these insights show how dynamic distributional modeling can complement historical analyses of top incomes by providing a structural interpretation of the mechanisms through which demographic forces reshape the entire income distribution.

Several directions for future research remain.
From a methodological perspective, improving the computational efficiency of the estimation procedure and developing more flexible state-space specifications would further enhance the applicability of the proposed framework.
From an empirical perspective, incorporating additional demographic and macroeconomic determinants would provide a richer understanding of the forces driving changes in income distributions.
Finally, applying the proposed methodology to other countries would help assess the generality of the framework and facilitate comparative analyses of distributional dynamics.
More broadly, the proposed framework provides a unified Bayesian approach for studying the evolution of income distributions from grouped data and offers a practical tool for model-based counterfactual analysis when individual-level microdata are unavailable.

\bibliographystyle{jae}
\bibliography{references}

\clearpage

\end{document}